\documentclass[journal]{IEEEtran}
\usepackage{cite}

\ifCLASSINFOpdf
   \usepackage[pdftex]{graphicx}
\else
\fi
\usepackage{amsmath}
\usepackage{url}

\usepackage{multirow,booktabs,colortbl,tabularx}

\usepackage{arydshln}

\usepackage{pifont}
\newcommand{\cmark}{\ding{51}}%
\newcommand{\xmark}{\ding{55}}%

\usepackage{xcolor,soul}

\begin{document}
%
\title{An Overview of Deep-Learning-Based Audio-Visual Speech Enhancement and Separation}
%
%
%

\author{Daniel~Michelsanti,~\IEEEmembership{Member,~IEEE,}
        Zheng-Hua~Tan,~\IEEEmembership{Senior~Member,~IEEE,}
        Shi-Xiong~Zhang,~\IEEEmembership{Member,~IEEE,}
        Yong~Xu,~\IEEEmembership{Member,~IEEE,}
        Meng~Yu,~\IEEEmembership{}
        Dong~Yu,~\IEEEmembership{Fellow~Member,~IEEE,}
        and~Jesper~Jensen,~\IEEEmembership{Member,~IEEE}

\thanks{ D. Michelsanti, and Z.-H. Tan are with the Department of Electronic Systems, Aalborg University, Aalborg 9220, Denmark (e-mail: \{danmi,zt\}@es.aau.dk).}
\thanks{ Shi-Xiong~Zhang, Yong~Xu, Meng~Yu, and~Dong~Yu are with Tencent AI Lab, Bellevue, WA, USA (e-mail: \{auszhang, lucayongxu, raymondmyu, dyu\}@tencent.com).}
\thanks{J. Jensen is with the Department of Electronic Systems, Aalborg University, Aalborg 9220, Denmark, and also with Oticon A/S, Sm{\o}rum 2765, Denmark (e-mail: jje@es.aau.dk).}
}

\maketitle

\begin{abstract}

\emph{Speech enhancement} and \emph{speech separation} are two related tasks, whose purpose is to extract either one or more target speech signals, respectively, from a mixture of sounds generated by several sources. Traditionally, these tasks have been tackled using signal processing and machine learning techniques applied to the available acoustic signals. {Since the visual aspect of speech is essentially unaffected by the acoustic environment, \emph{visual information} from the target speakers, such as lip movements and facial expressions, has also been used for speech enhancement and speech separation systems.} In order to efficiently fuse acoustic and visual information, researchers have exploited the flexibility of data-driven approaches, specifically \emph{deep learning}, achieving {strong} performance. The ceaseless proposal of a large number of techniques to extract features and fuse multimodal information has highlighted the need for an overview that comprehensively describes and discusses audio-visual speech enhancement and separation based on deep learning. In this paper, we provide a systematic survey of this research topic, focusing on the main elements that characterise the systems in the literature: \emph{acoustic features}; \emph{visual features}; \emph{deep learning methods}; \emph{fusion techniques}; \emph{training targets} and \emph{objective functions}. 
In addition, we review deep-learning-based methods for \emph{speech reconstruction from silent videos} and \emph{audio-visual sound source separation for non-speech signals}, since these methods can be more or less directly applied to audio-visual speech enhancement and separation.
 Finally, we survey commonly employed \emph{audio-visual speech datasets}, given their central role in the development of data-driven approaches, and \emph{evaluation methods}, because they are generally used to compare different systems and determine their performance.



\end{abstract}


%
\IEEEpeerreviewmaketitle

\section{Introduction}
%
%
%
%

\begin{figure*}
	\centering
		\includegraphics[scale=.22]{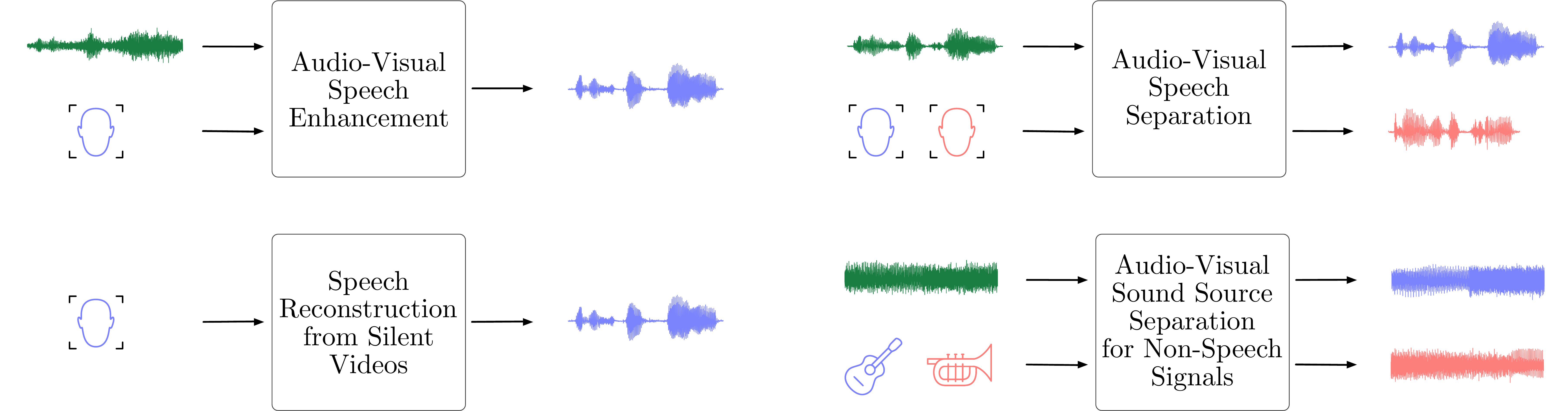}
	\caption{{Audio-visual sound source separation tasks. In audio-visual speech enhancement, the goal is to extract the target speech signal using a noisy observation of the target speech signal and visual information. Speech reconstruction from silent videos is a special case of audio-visual speech enhancement, where the noisy acoustic input signal is not provided. Audio-visual speech separation aims at extracting multiple target speech signals from a mixture and visual information of the target speakers. When the target sources are not speakers, but, for example, musical instruments, we refer to the task as audio-visual sound source separation for non-speech signals.}}
	\label{fig:tasks}
\end{figure*}

\IEEEPARstart{S}{peech} is one of the primary ways in which humans share information. A model that describes \emph{human speech communication} is the so-called \emph{speech chain}, which consists of two stages: \emph{speech production} and \emph{speech perception} \cite{deller2000discrete}. Speech production is the set of voluntary and involuntary actions that allow a person, i.e. a \emph{speaker}, to convert an idea expressed through a linguistic structure into a sound pressure wave. On the other hand, speech perception is the process happening mostly in the auditory system of a \emph{listener}, consisting of interpreting the sound pressure wave coming from the speaker. Some external factors, such as acoustic background noise,  can have an impact on the speech chain. Usually, normal-hearing listeners are able to focus on a specific acoustic stimulus, in our case the \emph{target speech} or \emph{speech of interest}, while filtering out other sounds \cite{bronkhorst2000cocktail, shinn2008selective}. This well-known phenomenon is called the \emph{cocktail party effect} \cite{cherry1953some}, because it resembles the situation occurring at a cocktail party.

Generally, the presence of high-level acoustic environmental noise or competing speakers poses several challenges to the speech communication effectiveness, especially for hearing-impaired listeners. Similarly, the performance of automatic speech recognition (ASR) systems can be severely impacted by a high level of acoustic noise. Therefore, several signal processing and machine learning techniques to be employed in e.g. hearing aids and ASR front-end units have been developed to perform \emph{speech enhancement}~(SE), which is the task of recovering the clean speech of a target speaker immersed in a noisy environment. {Especially when the receiver of an enhanced speech signal is a human, SE systems are often designed to improve two perceptual aspects: \emph{speech quality}, concerning how a speech signal sounds, and \emph{speech intelligibility}, concerning the linguistic content of a speech signal.} Some applications require the estimation of multiple target signals: this task is known in the literature as \emph{source separation} or \emph{speech separation} (SS), when the signals of interest are all speech signals.

Classical SE and SS approaches (cf. \cite{loizou2013speech, wang2006comp} and references therein) make assumptions regarding the statistical characteristics of the signals involved and aim at estimating the underlying target speech signal(s) according to mathematically tractable criteria. More recent methods based on \emph{deep learning} tend to depart from this \emph{knowledge-based} modelling, embracing a \emph{data-driven} paradigm. Most of these approaches treat SE and SS as supervised learning problems\footnote{{Sometimes, the approaches used in this context are more properly denoted as self-supervised or unsupervised learning techniques, since they do not use human-annotated datasets to learn representations of the data.}} \cite{wang2018supervised}.

The techniques mentioned above consider only acoustic signals, so we refer to them as audio-only SE (AO-SE) and audio-only SS (AO-SS) systems. However, speech perception is inherently multimodal, in particular audio-visual (AV), because in addition to the acoustic speech signal reaching the ears of the listeners, location and movements of some articulatory organs that contribute to speech production, e.g. tongue, teeth, lips, jaw and facial expressions, may also be visible to the receiver. Studies in neuroscience \cite{partan1999communication, golumbic2013visual} and speech perception \cite{sumby1954visual, mcgurk1976hearing} have shown that the visual aspect of speech has a potentially strong impact on the ability of humans to focus their auditory attention on a particular stimulus. {Even more importantly for SE and SS, visual information is immune to acoustic noise and competing speakers.} {This makes vision a reliable cue to exploit in challenging acoustic conditions.} These {considerations} inspired the first audio-visual SE (AV-SE) and audio-visual SS (AV-SS) works \cite{darrell2000audio, girin2001audio}, which demonstrated the benefit of using features extracted from the video of a speaker. Later, more complex frameworks based on classical statistical approaches have been proposed \cite{sodoyer2002separation, sodoyer2004developing, rivet2006mixing, rivet2007visual, maganti2007speech, almajai2010visually, naqvi2010multimodal, liang2012audio, naqvi2012multimodal, khan2013video, liu2013source, abel2014novel}, but they have very recently been outperformed by deep learning methods, such as \cite{wu2016multi, hou2017audio, gabbay2018visual, ephrat2018looking, afouras2018conversation, owens2018audio, lu2018listen, michelsanti2019deep, joze2019mmtm, lu2019audio, wu2019time, tan2019audio, sadeghi2019audio, gogate2019cochleanet, afouras2019my, gu2020multi, iuzzolino2020av, aldeneh2020self}. {In particular, deep learning allowed to overcome the limitations of knowledge-based approaches, making it possible to learn robust representations directly from the data and to jointly process AV signals with more flexibility.}

Despite the large amount of recent research and the interest in AV methods, no overview article currently focuses on deep-learning-based AV-SE and AV-SS. The survey article by Wang and Chen \cite{wang2018supervised} is the most extensive overview on deep-learning-based AO-SE and AO-SS for both single-microphone and multi-microphone settings, but it does not cover AV methods. The overview article by Rivet et al. \cite{rivet2014audiovisual} surveys AV-SS techniques, but it dates back to 2014, when deep learning was still not adopted for the task. Multimodal methods are also covered by Taha and Hussain \cite{ijca2018916290} in their survey on SE techniques. However, six AV-SE papers are discussed in total, and only one of these is based on deep learning. A limited number of deep learning approaches for AV-SE and AV-SS were described in \cite{rincon2019analysis, zhu2020deep}. In the first case, Rinc{\'o}n-Trujillo and C{\'o}rdova-Esparza \cite{rincon2019analysis} performed an analysis of deep-learning-based SS methods. They considered both AO-SS and AV-SS, with only five AV papers discussed. In the second case, Zhu et al. \cite{zhu2020deep} provided a bird's-eye view of several AV tasks, to which deep learning has been applied. 
Although AV-SE and AV-SS are discussed, the presentation covers only five approaches.

In this paper, we present an extensive survey of recent advances in AV methods for SE and SS, with a specific focus on deep-learning-based techniques. Our goal is to help the reader to navigate through the different approaches in the literature. Given this objective, we try not to recommend one approach over another based on its performance, because a comparison of systems designed for a heterogeneous set of applications might be unfair. Instead, we provide a systematic description of the main ideas and components that characterise deep-learning-based AV-SE and AV-SS systems, hoping to inspire and stimulate new research in the field. This is also the reason why current challenges and possible future directions are presented and discussed throughout the paper. Furthermore, {we provide} an overview of \emph{speech reconstruction from silent videos} and \emph{audio-visual sound source separation for non-speech signals} because they are strongly related to AV-SE and AV-SS (cf. Figure \ref{fig:tasks}). {Although other tasks may be considered related to AV-SE and AV-SS, their goal is substantially different. For example, AV speech recognition systems have some similarities with AV-SE and AV-SS, but they aim at finding the transcription of a video, not the clean target speech signal(s). We decide not to treat such methods in this overview.} Finally, we review AV datasets and evaluation methods, because they are two important elements used to train and assess the performance of the systems, respectively.

A list of resources for datasets, objective measures and several AV approaches can be accessed at the following link: \url{https://github.com/danmic/av-se}. There, we provide direct links to available demos and source codes, that would not be possible to include in this paper due to space limitations. Our goal is to allow both beginners and experts in the fields to easily access a collection of relevant resources.

{The rest of this paper is organised as follows.}
Section~\ref{subsec:prob_f} {presents the basic signal model to provide a formulation of the AV-SE and AV-SS problems.} 
Section~\ref{sec:av-se_ss} {introduces deep-learning-based AV-SE and AV-SS systems as a combination of several elements, described and discussed in the following sections, specifically:
acoustic features} (in Section~\ref{subsec:ac_f}); 
{visual features} (in Section~\ref{subsec:vis_f}); 
{deep learning methods} (in Section~\ref{subsec:deep_learn}); 
{fusion techniques} (in Section~\ref{subsec:fusion}); 
{training targets and objective functions} (in Section~\ref{subsec:tt_of}).
Afterwards, Section~\ref{sec:related} deals with speech reconstruction from silent videos and AV sound source separation for non-speech signals.
Section~\ref{sec:data} surveys relevant AV speech datasets that can be used to train deep-learning-based models. Section~\ref{sec:evaluation} presents a range of methodologies that may be considered  for performance assessment. Finally, Section \ref{sec:conclusion} provides a conclusion, summarising the principal concepts and the potential future research directions presented throughout the paper.


\section{Signal Model and Problem Formulation} \label{subsec:prob_f}

Let $h_s[n]$ denote the impulse response from the spatial position of the $s$-th target source to the microphone, with $n$ indicating a discrete-time index. Furthermore, let $h_s[n]=h_s^{e}[n]+h_s^{l}[n]$, where $h_s^{e}[n]$ is the early part of $h_s[n]$ (containing the direct sound and low-order reflections) and $h_s^{l}[n]$ is the late part of $h_s[n]$. Assuming a total number of $S$ target speech signals and a number of $C$ additive noise sources, the observed acoustic mixture signal can be modelled as:
\begin{align}
	\label{eq:sig_mod_1_3} y[n]     = {\sum^{S}_{s=1}{x_s[n]} + d[n]}
\end{align}
with:
\begin{align}
	\label{eq:sig_mod_1_1} x_s[n] &= {x'_s[n]*h_s^{e}[n]}, \\
	\label{eq:sig_mod_1_2} d[n]     &= {\sum^{S}_{s=1}{ x'_s[n]*h_s^{l}[n]} + \sum^{C}_{c=1}{ d_c[n]}},
\end{align}
\noindent where ${x'_s[n]}$ is the speech signal emitted at the $s$-th target speaker position, ${x_s[n]}$ is the clean speech signal from the $s$-th target speaker at the microphone (including low-order reflections),  $d_c[n]$ is the signal from the $c$-th noise source as observed at the microphone and $d[n]$ indicates the total contribution from noise and late reverberations. Furthermore, let $v[m]$ indicate the observed two-dimensional visual signal, with $m$ denoting a discrete-time index different from $n$, because the acoustic and the visual signals are usually not sampled with the same sampling rate.


Given $y[n]$ and $v[m]$, the task of AV-SS consists of determining estimates $\hat{x}_s[n]$ of ${x}_s[n]$\footnote{While preserving early reflections is important in some applications (e.g. hearing aids), in other cases the goal is to determine only estimates of $x'_s[n]$. This observation does not have a big impact on the formulation of the problem, therefore we are not going to make a distinction between the two cases.}, with $s = 1, \dots, S$. In some setups, additional information is available, for example a speakers' enrolment acoustic signal and a training set collected under time and location different from the recordings of $y[n]$ and $v[m]$.

When $S = 1$, we refer to the task as AV-SE and rewrite Eq.~(\ref{eq:sig_mod_1_3}) as: 
\begin{equation}
	y[n] = x[n] + d[n],
\label{eq:sig_mod_2}
\end{equation}
\noindent with $x[n]$ denoting $x_1[n]$.

Due to the linearity of the short-time Fourier transform (STFT), it is possible to express the acoustic signal model of Eqs.~(\ref{eq:sig_mod_1_3}) and (\ref{eq:sig_mod_2}) in the time-frequency (TF) domain as: 
\begin{equation}
	Y(k,l) = \sum^{S}_{s=1}{X_s(k,l)} + D(k,l) ,
\label{eq:sig_mod_1TF}
\end{equation}
\noindent for SS, and as:
\begin{equation}
	Y(k,l) = X(k,l) + D(k,l) ,
\label{eq:sig_mod_2TF}
\end{equation}
\noindent for SE, where $k$ denotes a frequency bin index, $l$ indicates a time frame index, and $Y (k, l)$, $X_s(k, l)$ and $D(k, l)$ are the short-time Fourier transform (STFT) coefficients of the mixture, the $s$-th target signal, and the noise, respectively.

The definitions provided above are valid for single-microphone single-camera AV-SE and AV-SS. It is possible to extend all the concepts to the case of multiple acoustic and visual signals. Let $F$ and $P$ be the number of cameras and microphones of a system, respectively. We denote as $v_f[m]$ the observed visual signal with the $f$-th camera. Assuming S speakers to separate, then the acoustic mixture as received by the $p$-th microphone can be modelled as: 
\begin{align}
	\label{eq:multi_ch_3} y_p[n]      = {\sum^{S}_{s=1}{x_{ps}} + d_p[n]}.
\end{align}
with:
\begin{align}
	\label{eq:multi_ch_1} x_{ps}[n] & = {x'_s[n] * h^e_{ps}[n]} \\
	\label{eq:multi_ch_2} d_p[n]     & = {\sum^{S}_{s=1}{ x'_s[n]*h_{ps}^{l}[n]} + \sum^{C}_{c=1}{ d_{pc}[n]}}
\end{align}
%
 \noindent In this case, the SS task consists of determining estimates $\hat{x}_s[n]$ of ${x}_{p^*s}[n]$ for $s = 1, \dots, S$, given $v_f[m]$ with $f = 1, \dots, F$, $y_p[n]$ with $p = 1, \dots, P$ and any other additional information, assuming that the microphone with index $p=p^*$ is a pre-defined reference microphone.

\section{Audio-Visual Speech Enhancement and Separation Systems} \label{sec:av-se_ss}



The problems of AV-SE and AV-SS have recently been tackled with supervised learning techniques, specifically deep learning methods. Supervised deep-learning-based models can automatically learn how to perform SE or SS after a training procedure, in which pairs of degraded and clean speech signals, together with the video of the speakers, are presented to them. Ideally, deep-learning-based systems should be trained using data that is representative of the settings in which they are deployed. This means that in order to have good performance in a wide variety of settings, very large AV datasets for training and testing need to be collected. In practice, the systems are trained using a large number of complex acoustic scenes that are synthetically generated {using a mix-and-separate paradigm} \cite{zhao2018sound}{, where target speech signals are added to signals from sources of interference at several signal to noise ratios (SNRs)}. This way of generating synthetic training material has empirically shown its effectiveness in both audio-only (AO) and AV settings, since speech signals processed with systems trained in this way improve in terms of both estimated speech quality and intelligibility \cite{kolbaek2017speech, yu2017permutation, ephrat2018looking, afouras2018conversation}.

\begin{figure}
	\centering
		\includegraphics[scale=.2]{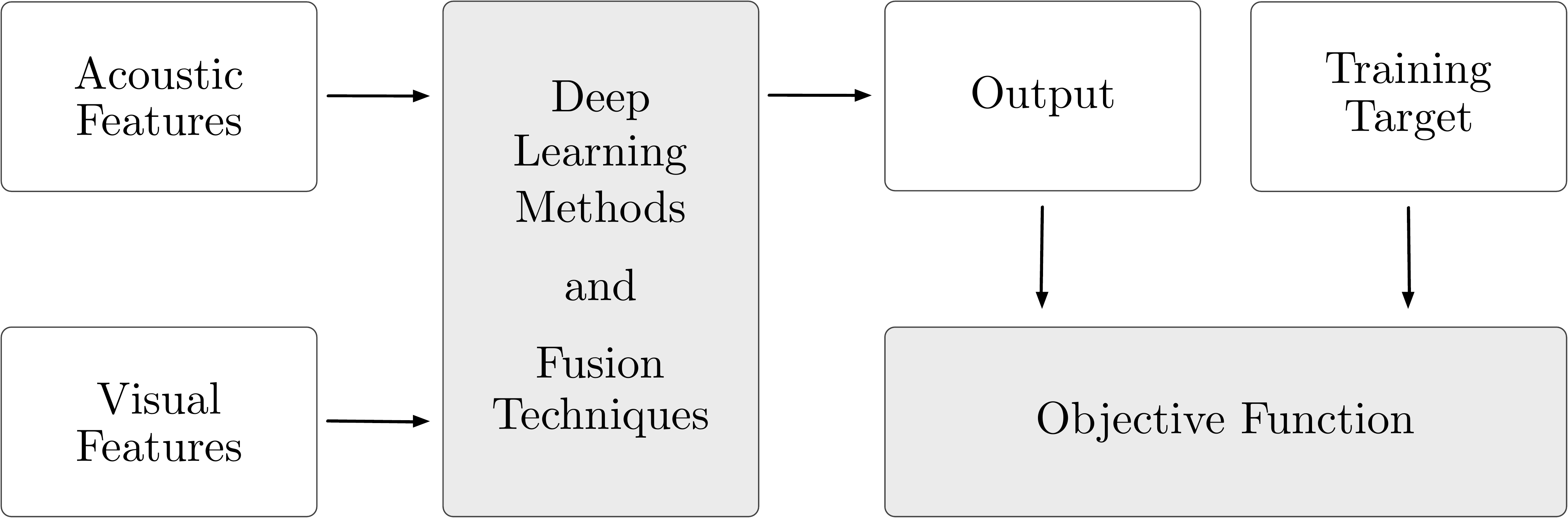}
	\caption{Interconnections between the main elements of a generic audio-visual speech enhancement/separation system based on deep learning. White boxes represent data, while grey boxes represent processing blocks.}
	\label{fig:avss_characteristics}
\end{figure}

{In the following sections, we focus on the main elements of deep-learning-based AV-SE and AV-SS systems, i.e.}: acoustic features; visual features; deep learning methods; fusion techniques; training targets and objective functions\footnote{Training targets and objective functions are not used during inference.}. Figure~\ref{fig:avss_characteristics} provides a conceptual block diagram illustrating the interconnections of these elements. 

%
%
%

\section{{Acoustic Features}} \label{subsec:ac_f}

{As represented in Figure} \ref{fig:avss_characteristics}, {acoustic features are one of the main elements of AV-SE and AV-SS systems. In this Section, we report which features are used in the literature, following the list provided in Table} \ref{tab:avss_acoustic_feat}.

\subsection{{Single-Microphone Features}}

AV-SE and AV-SS systems process acoustic information (cf. Figure \ref{fig:avss_characteristics}). As can be seen in Table \ref{tab:avss_acoustic_feat}, the predominant acoustic input feature is the (potentially transformed) magnitude spectrogram of a single-microphone recording, sometimes in the log mel domain, like in \cite{gabbay2018visual}. However, a magnitude spectrogram is generally an incomplete representation of the acoustic signal, because it is computed from STFT coefficients which are complex-valued. Recent works have used as acoustic input to the AV system either the magnitude spectrogram and the respective phase \cite{afouras2018conversation, afouras2019my, 9054180}, the real and the imaginary parts of the complex spectrogram \cite{ephrat2018looking, sun2020attention, luo2019audio, inan2019evaluating, ideli2019multi}, or directly the raw waveform \cite{wu2019time, ideli2019visually}. Although these approaches allow to incorporate and process the full information of an acoustic signal, research in this area is still active and suggests that there is still room for improvement by exploiting the full information of the noisy speech signal~\cite{luo2019conv, yin2020phasen}.

\subsection{{Speaker Embeddings}}

Since Wang et al. \cite{wang2018voicefilter} showed that an AO system can successfully extract the speech of interest from a mixture signal when conditioned on the \emph{speaker embedding} vector of an enrolment audio signal of the target spreaker, several AV-SE and AV-SS systems have made use of a similar idea. Luo et al. \cite{luo2019audio} showed that i-vectors \cite{dehak2010front}, a low-dimensional representation of a speech signal effective in speaker verification, recognition and diarisation \cite{verma2015vectors}, were particularly effective for AV-SS of same gender speakers, obtaining a large improvement over an AV baseline model that did not incorporate speaker embeddings. Afouras et al. \cite{afouras2019my} extracted a compact speaker representation from an enrolment speech signal  with the deep-learning-based method in \cite{xie2019utterance} and obtained good performance for mixtures of two and three speakers, especially when face occlusions occurred. In addition, their system could learn the speaker representation on the fly by using the enhanced magnitude spectrogram obtained from a first run of the algorithm without speaker embedding. This essentially bypassed the need for enrolment audio, which is cumbersome or even impossible to collect in certain applications. The approach in \cite{gu2020multi} also used a pre-trained deep-learning-based model \cite{zhang2018text} to extract a speaker representation from an additional audio recording. The results indicate that visual information of the speaker's lips is more important than the information contained in the speaker embedding vector, and that their combination led to a general performance improvement. Instead of adopting a pre-trained model, Ochiai et al. \cite{ochiai2019multimodal} decided to use a sequence summarising neural network (SSNN) \cite{vesely2016sequence}, which was jointly trained with the main separation model. Their experiments showed that similar outcomes could be obtained when the enrolment audio and the visual information were used as input in isolation, but better performance was achieved when used at the same time. 
In general, all these approaches show that speaker embeddings, when extracted from an available additional speech utterance from the target speaker, can be useful, confirming the results obtained in the AO domain  \cite{wang2018voicefilter}.

\begin{table}
\caption{{List of acoustic features in audio-visual speech enhancement and separation papers.}}
\centering
\resizebox{0.48\textwidth}{!}{%
\begin{tabular}{l l}
\toprule
 Acoustic Features & AV-SE/SS papers \\
\midrule
 Magnitude spectrogram & \cite{ adeeltowards, adeel2019lip, adeel2019contextual, adeel2019novel, afouras2018conversation, afouras2019my, aldeneh2020self, arriandiaga2019audio, chuang2020lite, chung2020facefilter, gabbay2018visual, gabbay2018seeing, gogate2018dnn,  gogate2019cochleanet}\\
& \cite{gu2020multi, hou2017audio, hou2016audio, ideli2019multi, iuzzolino2020av, joze2019mmtm, khan2018using, li2020visual, 9054180}\\
& \cite{lu2018listen, lu2019audio, michelsanti2019training, michelsanti2019effects, michelsanti2019deep, morrone2019face, ochiai2019multimodal,  owens2018audio}\\ 
& \cite{pasa2019joined, qu2020multimodal, sadeghi2019mixture, sadeghi2019audio, sadeghi2019robust, tan2019audio, 9053033, wu2016multi, xu2020neural}\\
 Phase$^a$ \rule{0pt}{2.5ex} & \cite{afouras2018conversation, afouras2019my, 9054180} \\
 Complex spectrogram \rule{0pt}{2.5ex} &  \cite{ephrat2018looking, sun2020attention, luo2019audio, inan2019evaluating, ideli2019multi} \\
 Raw waveform \rule{0pt}{2.5ex} & \cite{wu2019time, ideli2019visually}\\
 Speaker embeddings  \rule{0pt}{2.5ex} & \cite{ochiai2019multimodal, afouras2019my, luo2019audio, gu2020multi, qu2020multimodal}\\
 IPD \(|\) cosIPD \(|\) sinIPD \rule{0pt}{2.5ex}& \cite{gu2020multi, ideli2019multi, xu2020neural} \: \(|\) \: \cite{tan2019audio, ideli2019multi} \: \(|\) \: \cite{ideli2019multi}\\
 Angle feature \rule{0pt}{2.5ex} &\cite{tan2019audio, gu2020multi, xu2020neural} \\
\bottomrule
 \multicolumn{2}{l}{$^a$Only if it is used in processing, not just to reconstruct the signal. \rule{0pt}{2.5ex}}\\
\end{tabular}}
\label{tab:avss_acoustic_feat}
\end{table}

\subsection{{Multi-Microphone Features}}

The spatial information contained in multi-channel acoustic recordings provides an informative cue complementary to spectral information for separating multiple speakers. Specifically, inter-channel phase differences (IPDs) \cite{gu2019end}, inter-channel time differences (ITDs) \cite{jiang2014binaural}, inter-channel level differences (ILDs) \cite{jiang2014binaural}, directional statistics \cite{chen2018multi} or simply mixture STFT vectors \cite{ochiai2017does} are used in multi-channel deep-learning-based systems to perform SE or SS. Among these features, IPDs are widely applied due to their robustness to reverberation and microphone sensitivities \cite{gu2020multi}. However, because of the well known issues of spatial aliasing and phase wrapping, IPDs can be the same even for spatially separated sources with different time delays in particular frequencies. This causes fundamental difficulties in separating one source from another. Wang et al. \cite{wang2018multi} proposed to concatenate cosine IPDs (cosIPDs) and sine IPDs (sinIPDs) with log magnitudes as input of their AO system. With this strategy, spectral features can help to resolve the IPDs ambiguity. In addition, the combination of cosIPDs and sinIPDs is preferred over IPDs, because it exhibits a continuous helix structure along frequency due to the Euler formula \cite{wang2020deep}, while IPDs suffer from abrupt discontinuities caused by phase wrapping. In AV-SE and AV-SS, systems used IPDs \cite{gu2020multi}, cosIPDs \cite{ideli2019multi, tan2019audio} and sinIPDs~\cite{ideli2019multi}. Some AV multi-microphone approaches \cite{gu2020multi,tan2019audio} effectively included also an angle feature \cite{chen2018multi}, which computes the averaged cosine distance between the target speaker steering vector and IPD on all selected microphone pairs.

\subsection{{Shortcomings and Future Research}}

{As reported above, the vast majority of AV-SE and AV-SS systems use a TF representation of a single-channel acoustic signal as acoustic features. Although a limited number of AV approaches adopt a time-domain signal} \cite{wu2019time, ideli2019visually} {or multi-microphone cues} \cite{gu2020multi, ideli2019multi, tan2019audio, xu2020neural} {as acoustic features, there is still room to explore these aspects in future research.  In particular, the integration of multi-microphone features with visual information still needs to be investigated further, for example in order to correctly estimate the direction of arrival of the target speech which is hard at low SNRs.}

\section{{Visual Features}} \label{subsec:vis_f}

{Besides acoustic information, AV-SE and AV-SS systems also exploit visual information.} In general, the use of {vision} allows AV systems to obtain a performance improvement over AO systems. A more detailed analysis regarding the actual contribution of vision for AV-SE was conducted in \cite{aldeneh2020self}. In particular, visual features were shown to be important to get not only high-level information about speech and silence regions of an utterance, but also fine-grained information about articulation. Although improvements were shown for all \emph{visemes}\footnote{A viseme is the basic unit of visual speech and represents what a phoneme is for acoustic speech \cite{massaro2014speech}.}, sounds that are easier to distinguish visually were the ones that improved the most with an AV-SE system.

{The focus of this Section is on visual features, following the list in Table} \ref{tab:avss_visual_feat}. {Before talking about visual features, we provide information about \emph{face detection} and \emph{tracking}, because a solution of these problems is critical in AV-SE and AV-SS systems.}


\subsection{{Face Detection and Tracking}}

Given a video recording, the first step of most AV-SE and AV-SS systems is to determine the number of speakers in it and track their faces across the visual frames.  This is usually performed by face detection \cite{viola2004robust, dlib09, liu2016ssd} and tracking \cite{lucas1981iterative, tomasi1991detection} algorithms. 
This approach allows to considerably reduce the dimensionality of the input and, as a consequence, the number of parameters of the SE and the SS models, because only crops of the target faces are considered.
In addition, face detection is one way to determine the number of speakers in a scene, {an information that can be used by the SS systems that can handle only a fixed number of target speech signals} (e.g. \cite{ephrat2018looking}), because a priori knowledge of the number of speakers is needed to choose a specific trained multi-speaker model. From these considerations, we can understand the critical importance of face detection and tracking algorithms: if they fail, all the later modules would fail as well. Therefore, robust face tracking, in particular under varying light conditions, occlusions etc. is essential to guarantee high performance in real-world scenarios.

\subsection{{Raw Visual Data}}

Once that the video frames of the speaker's face are available, visual features can be used by AV-SE and AV-SS approaches (cf. Table \ref{tab:avss_visual_feat}). Many systems, such as \cite{gabbay2018seeing} and \cite{gabbay2018visual}, directly use a crop around the face or the mouth of the target speaker(s) as input, {sometimes aligned using an affine transformation} \cite{iuzzolino2020av}.
This approach is not always convenient: learning to perform a task from high-dimensional input consisting of raw pixels with a neural network is usually challenging and requires a large amount of data \cite{inan2019evaluating, luo2019audio}. Hence, several approaches are employed to reduce the input dimensions by extracting different types of features from the raw pixel input, as we report in the following.

\begin{table}
\caption{{List of visual features in audio-visual speech enhancement and separation papers.}}
\centering
\resizebox{0.48\textwidth}{!}{%
\begin{tabular}{l l}
\toprule
 Visual Features & AV-SE/SS papers \\
\midrule
 Raw pixels: \\
 \quad - Mouth \rule{0pt}{2.5ex}& \cite{aldeneh2020self, gabbay2018visual, gogate2018dnn, gogate2019cochleanet, gu2020multi, hou2017audio, iuzzolino2020av, joze2019mmtm, lu2018listen}\\ &\cite{lu2019audio, michelsanti2019training, michelsanti2019effects, michelsanti2019deep, wu2016multi, tan2019audio, sadeghi2019audio, 9053033}\\
&\cite{sadeghi2019robust, sadeghi2019mixture, xu2020neural} \\
 \quad - Face \rule{0pt}{2.5ex} & \cite{gabbay2018seeing}\\
 AAM of mouth region \rule{0pt}{2.5ex} & \cite{khan2018using}\\
 2D-DCT of mouth region \rule{0pt}{2.5ex} & \cite{adeeltowards, adeel2019lip, adeel2019contextual, adeel2019novel} \\
 Optical flow \rule{0pt}{2.5ex}& \cite{gabbay2018seeing, lu2018listen, lu2019audio, li2020visual,arriandiaga2019audio} \\
 Landmark-based features \rule{0pt}{2.5ex} & \cite{hou2016audio, morrone2019face, pasa2019joined, li2020visual}\\
 Multisensory features \rule{0pt}{2.5ex} & \cite{owens2018audio}\\
 Face recognition embedding \rule{0pt}{2.5ex} & \cite{ephrat2018looking, ochiai2019multimodal, sun2020attention, luo2019audio, inan2019evaluating} \\
 VSR embedding \rule{0pt}{2.5ex} & \cite{wu2019time, afouras2018conversation, inan2019evaluating, ideli2019visually, ideli2019multi, sadeghi2019audio, 9054180, afouras2019my} \\
 Facial appearance embedding \rule{0pt}{2.5ex} & \cite{chung2020facefilter, qu2020multimodal}\\
 Compressed mouth frames \rule{0pt}{2.5ex} & \cite{chuang2020lite} \\
 Speaker direction \rule{0pt}{2.5ex} & \cite{gu2020multi, tan2019audio, xu2020neural} \\
\bottomrule
\end{tabular}}
\label{tab:avss_visual_feat}
\end{table}

\subsection{{Low-Dimensional Visual Features}}

Khan et al. \cite{khan2018using} reduced the dimensionality of the visual information with an active appearance model (AAM) \cite{cootes2001active}, which is a framework that combines appearance-based and shape-based features through principal component analysis (PCA). Other classical approaches have also been used for visual feature extraction. For example, some works \cite{adeeltowards, adeel2019lip, adeel2019contextual, adeel2019novel} produced a vector of pixel intensities from the lip region of the speaker with a 2-D discrete cosine transform (DCT). Alternatively, optical flow features were used as an additional input in \cite{gabbay2018seeing, lu2018listen, lu2019audio, li2020visual} to explicitly incorporate the motion information in the system.

Research has also been conducted to investigate the use of \emph{facial landmark points}. Hou et al. \cite{hou2016audio} considered a representation of the speaker's mouth consisting of the coordinates of 18 points. Distances for each pair of these points were computed and the 20 elements with the highest variance across an utterance were provided to the SE network. Instead of the distance for each pair of landmark points, Morrone et al. \cite{morrone2019face} obtained a differential motion feature vector by subtracting the face landmark points of a video frame with the points extracted from the previous frame. Motion of landmarks points was also exploited by Li et al. \cite{li2020visual}, who first computed the distance for every symmetric pair of lip landmark points in the vertical and the horizontal directions, and then defined a variation vector of the lip movements consisting of the differences between the distance vectors of two contiguous video frames. This distance-based motion vector was finally combined with aspect ratio features. 

A different approach consists of extracting embeddings, i.e. meaningful representations in a typically low dimensional projected space, with a neural network pre-trained on a related task. For example, Owens and Efros \cite{owens2018audio} proposed to use multisensory features. They designed a deep-learning-based system that could recognise whether the audio and the video streams of a recording were synchronised. The features extracted from such a network provided an AV representation that allows to achieve superior performance compared to an AO-SE approach. Besides multisensory features, embeddings extracted with {models trained on face recognition} \cite{ephrat2018looking} {or visual speech recognition (VSR)} \cite{afouras2018conversation} {tasks} have been shown to be effective. {\.I}nan et al. \cite{inan2019evaluating} performed a study to evaluate the differences between these two kinds of embeddings. Their results showed that VSR embeddings were able to separate voice activity and silence regions better than face recognition embeddings, which could provide a better distinction between speakers instead. Overall, the performance obtained with VSR embeddings was superior, because they allowed to easier characterise lip movements. Another study \cite{wu2019time} further investigated VSR embeddings, showing that the use of features extracted with a model trained for phone-level classification led to better results if compared to the adoption of word-level embeddings.

\subsection{{Still Images as Visual Input}}

Attempts \cite{chung2020facefilter, qu2020multimodal} have been made to exploit the information of a still image of the target speaker instead of a video. This approach outperformed a system that used only the audio signals, because there exists a cross-modal relationship between the voice characteristics of a speaker and their facial appearance \cite{kim2018learning, oh2019speech2face}. This explains why facial features can guide the extraction of the target speech from a mixture. The advantage of using a still image is the reduced complexity of the overall system, although the dynamic information of the video is lost, {limiting the system performance considerably.}

\subsection{{Visual Information in Multi-Microphone Approaches}}

When the information from \emph{multiple microphones} is available, the location of the target speaker with respect to the microphone array can be used for spatial filtering, i.e. beamforming. In \cite{gu2020multi, tan2019audio}, the target direction is estimated with a face detection method. In more complicated scenarios, where people move and turn their heads, face detection might fail over several visual frames. The use of features from the speaker's body might help in building a more robust target source tracker.

\subsection{{Shortcomings and Future Research}}

{Current AV-SE and AV-SS approaches only process the visual signal from a single camera. However, previous research on VSR} \cite{kumar2007profile, lan2012view} {showed that the use of a speaker's profile view can outperform the frontal view. We expect that combining the information from several cameras to capture the different views of a talking face could improve current AV-SE and AV-SS systems. Multi-view input signals were used in approaches for speech reconstruction from silent videos and are reported in Section} \ref{sec:related}.

{Other} future challenges include the extraction of features with low complexity algorithms that can be robust to illumination changes, occlusion and pose variations. At the moment, these robustness issues are tackled with a {noise-aware training, where} the data is artificially modified to include such perturbations \cite{afouras2019my}. New opportunities to build low-latency systems that are energy-efficient and robust to light changes are given by \emph{event cameras}. In contrast to conventional frame-based cameras, event cameras are asynchronous sensors that output changes in brightness for each pixel only when they occur. They have low latency, high dynamic range and very low power consumption \cite{lichtsteiner2008128}. Arriandiaga et al. \cite{arriandiaga2019audio} showed that the SE results obtained with optical flow features, extracted from an event camera, are on par with a frame-based approach. The main limitation of exploiting the full potential of event cameras is that existing image processing algorithms cannot be employed, due to the inherently different nature of the data produced by them. Research in this area is expected to bring novel algorithms and performance improvements.


\section{{Deep Learning Methods}} \label{subsec:deep_learn}

As illustrated in Figure \ref{fig:avss_characteristics}, after the feature extraction stage, the actual processing and fusion of acoustic and visual information is performed with a combination of deep neural network models. {The main advantage of using these models instead of knowledge-based techniques is the possibility to learn representations of the acoustic and visual modalities at several levels of abstraction and flexibly combine them.} Although a detailed exposition of general deep learning architectures and concepts \cite{goodfellow2016deep} is outside of the scope of this paper, {in this Section, we provide a brief presentation of the deep neural network models used in AV-SE and AV-SS systems, as listed in Table} \ref{tab:avss_deep_methods}.

\begin{table}
\caption{{List of deep learning methods in audio-visual speech enhancement and separation papers.}}
\centering
\resizebox{0.48\textwidth}{!}{%
\begin{tabular}{l l}
\toprule
Deep Learning Methods & AV-SE/SS papers \\
\midrule
 FFNN & \cite{adeel2019contextual, aldeneh2020self, afouras2019my, adeeltowards, adeel2019lip, adeel2019novel, chuang2020lite, chung2020facefilter, ephrat2018looking, gabbay2018visual, gabbay2018seeing, gogate2018dnn,  gogate2019cochleanet}\\ 
 &\cite{hou2017audio, hou2016audio, ideli2019multi, inan2019evaluating, khan2018using, li2020visual, lu2018listen, lu2019audio, 9054180}\\
& \cite{luo2019audio, michelsanti2019training, michelsanti2019effects, michelsanti2019deep, morrone2019face, ochiai2019multimodal, qu2020multimodal, sadeghi2019audio, sadeghi2019robust, sadeghi2019mixture}\\
& \cite{sun2020attention, 9053033, wu2016multi, tan2019audio} \\
 CNN \rule{0pt}{2.5ex} & \cite{afouras2018conversation, afouras2019my, aldeneh2020self, adeel2019novel, adeel2019contextual, chuang2020lite, chung2020facefilter, ephrat2018looking, gabbay2018visual, gogate2018dnn,  gogate2019cochleanet}\\ 
 &\cite{gu2020multi, hou2017audio, inan2019evaluating, ideli2019visually, iuzzolino2020av, ideli2019multi, li2020visual, 9054180, lu2018listen, lu2019audio}\\
&\cite{luo2019audio, michelsanti2019training, michelsanti2019effects, michelsanti2019deep, ochiai2019multimodal, owens2018audio, qu2020multimodal, sun2020attention, tan2019audio}\\
&\cite{9053033, wu2019time, wu2016multi, xu2020neural} \\
 AE \rule{0pt}{2.5ex} & \cite{chuang2020lite, iuzzolino2020av, ideli2019multi, michelsanti2019training, michelsanti2019effects, michelsanti2019deep, owens2018audio, joze2019mmtm, gabbay2018visual}\\
 & \cite{sadeghi2019audio, sadeghi2019robust, sadeghi2019mixture}\\
 LSTM \rule{0pt}{2.5ex} & \cite{aldeneh2020self, chuang2020lite, 9053033, inan2019evaluating, joze2019mmtm, adeeltowards, adeel2019lip, gogate2018dnn, adeel2019contextual, gogate2019cochleanet, adeel2019novel}\\
 BiLSTM \rule{0pt}{2.5ex} & \cite{li2020visual, iuzzolino2020av, ideli2019multi,  arriandiaga2019audio,  afouras2019my, luo2019audio,   lu2018listen, lu2019audio, ephrat2018looking}\\ 
 & \cite{morrone2019face, ochiai2019multimodal, pasa2019joined, qu2020multimodal, sun2020attention, tan2019audio, wu2016multi}\\
 Skip connections \rule{0pt}{2.5ex} & \cite{michelsanti2019training, michelsanti2019effects, michelsanti2019deep, owens2018audio, iuzzolino2020av, ideli2019multi, joze2019mmtm} \\
 Residual connections \rule{0pt}{2.5ex} & \cite{gu2020multi, chung2020facefilter, iuzzolino2020av, ideli2019multi, afouras2018conversation, afouras2019my, joze2019mmtm, ideli2019visually, gabbay2018seeing, 9054180}\\
 &\cite{wu2019time, tan2019audio, xu2020neural} \\
\bottomrule
\end{tabular}}
\label{tab:avss_deep_methods}
\end{table}

\subsection{{Feedforward Neural Networks}}

One of the most used architectures is the feedforward fully-connected neural network (FFNN), also known as multilayer perceptron (MLP). A FFNN consists of several artificial neurons, or \emph{nodes}, organised into a number of \emph{layers}. The network is fully-connected because each node shares a connection with every node belonging to the previous layer. In addition, it is feedforward since the information flows only in one direction from the input layer to the output layer, through the intermediate layers, called \emph{hidden layers}. In order to act as a universal approximator \cite{hornik1989multilayer, cybenko1989approximation, hornik1991approximation}, i.e. being able to approximate arbitrarily well any function which maps intervals of real numbers to some real interval, a FFNN needs also to include activation functions, like sigmoid or ReLU, which allow to model potential non-linearities of the function to approximate.

Another kind of feedforward network is the convolutional neural network (CNN) \cite{lecun1989generalization}. While in FFNNs each node is connected with all the nodes of the previous layer, CNNs are based on the \emph{convolution operation}, which leverages \emph{sparse connectivity}, \emph{parameter sharing} and \emph{equivariance to translation} \cite{goodfellow2016deep}. Sometimes, a convolutional layer is followed by a \emph{pooling operation}, which performs a downsampling, for example by local maximisation, to reduce the amount of parameters and obtain \emph{invariance to local transformations}. In AV-SE and AV-SS systems, CNNs are generally used to process the visual frames and automatically extract visual features \cite{wu2016multi}. They are also adopted for the acoustic signals, to process either the spectrogram \cite{gabbay2018visual} or the raw waveform \cite{wu2019time}. Since in SE and SS the acoustic input and the output shares a similar structure, some approaches, such as \cite{iuzzolino2020av, michelsanti2019training, owens2018audio}, adopted a convolutional autoencoder (AE) architecture, sometimes including skip-connections like in U-Net \cite{ronneberger2015u} to allow the information to flow despite the bottleneck.

The training of feedforward neural networks, i.e. the update of the network parameters, is performed e.g. using stochastic gradient descent (SGD) \cite{robbins1951stochastic, kiefer1952stochastic} to minimise an objective function (see Section \ref{subsec:tt_of} for further details) using the backpropagation algorithm \cite{rumelhart1986learning} for gradient computation. Variations of SGD are also adopted, in particular RmsProp \cite{Tieleman2012} and Adam \cite{kingma2014adam}. Although increasing the number of hidden layers, i.e. the network \emph{depth}, usually leads to a performance increase \cite{simonyan2014very}, two issues often arise: \emph{vanishing/exploding gradient} \cite{bengio1994learning, glorot2010understanding} and \emph{degradation problem} \cite{he2016deep}. These issues are generally addressed with \emph{batch normalisation} \cite{ioffe2015batch} and \emph{residual connections} \cite{he2016deep}, respectively, both extensively adopted in AV-SE and AV-SS systems.

\subsection{{Recurrent Neural Networks}}

When dealing with speech signals, a different family of neural networks is also used: recurrent neural networks (RNNs) \cite{rumelhart1986learning}. The reason is that RNNs were designed to process sequential data. Therefore, they are particularly suitable for speech signals, in which the temporal dimension is important. The training of RNNs is performed with backpropagation through time \cite{werbos1990backpropagation} and, similarly to feedforward neural networks, vanishing/exploding gradient issues are common. The most effective solution to the problem is to introduce paths in which the gradient could flow through time and regulate the propagation of information with \emph{gates}. This class of networks are called gated RNNs, and among them the most adopted are long short-term memory (LSTM) \cite{hochreiter1997long, felix2000learning} and gated recurrent unit (GRU) \cite{cho2014learning}. Although these models have a causal structure, architectures in which the output at a given time step depends on the whole sequence, including past and future observations, are also common, and they are known as bidirectional RNNs (BiRNNs) \cite{schuster1997bidirectional}, bidirectional LSTMs (BiLSTMs) and bidirectional GRUs (BiGRUs).  


\subsection{{Shortcomings and Future Research}}

Compared to knowledge-based approaches, deep learning methods have some disadvantages that we expect to be addressed in future works. First of all, neural network architectures need to be trained with a large amount of data to generalise well to a wide variety of speakers, languages, noise types, SNRs, illumination conditions and face poses. A big step in the evolution of AV-SE and AV-SS systems occurred when researchers started to train the models with large-scale AV datasets \cite{ephrat2018looking, owens2018audio, afouras2018conversation}. An interesting research direction would be to study the possibility of training deep-learning-based systems with a smaller amount of data without degrading the performance in unknown scenarios \cite{gogate2018dnn, gogate2019cochleanet}. In this context, it would be relevant to explore unsupervised learning techniques, such as the one proposed by Sadeghi~et~al. \cite{sadeghi2019audio, sadeghi2019robust, sadeghi2019mixture}, who extended a previous work on AO-SE \cite{leglaive2018variance} and adopted variational auto-encoders (VAEs) for AV-SE. In their approach, there is no need of mixing many different noise types with the speech of interest at several SNRs, because the system models directly the clean speech. Despite this attempt, a supervised learning approach that learns a mapping from noisy to clean speech or from a mixture to separated speech signals is still the preferred way to tackle AV-SE and AV-SS, because it allows to reach state-of-the-art performance. 

Furthermore, typical paradigms employed for training AV-SE and AV-SS systems assume that the sound sources of a scene are independent from each other. This assumption is adopted for convenience, because collecting actual speech in noise data is costly. However, it is often wrong, since speakers tend to change the way they speak, when they are immersed in a noisy environment, in order to make their speech more intelligible. This phenomenon is known in the literature as \emph{Lombard effect} \cite{lombard1911signe, brumm2011evolution}. 
Recent work \cite{michelsanti2019effects, michelsanti2019deep} investigated the impact of this effect on data-driven AV-SE models, showing that training a system with Lombard speech is beneficial especially at low SNRs. Therefore, the performance of most deep-learning-based AV-SE systems is affected by the fact that data used for training does not match real conditions.

Another issue especially for low-resource devices is that deep learning models are usually computationally expensive, because data needs to be processed with an algorithm consisting of millions of parameters in order to achieve satisfactory performance. It is important to explore novel ways to reduce the model complexity without reducing the speech quality and intelligibility of the processed signals.

\section{{Fusion Techniques}} \label{subsec:fusion}

As previously mentioned, AV-SE and AV-SS systems typically consist of a combination of the neural network architectures presented above, which allows to fuse the acoustic and visual information in several ways. {In this Section, we present several fusion strategies used in the literature. However, we first introduce the problem of \emph{AV synchronisation}, which is relevant when acoustic and visual data need to be integrated.}

\begin{figure}
	\centering
		\includegraphics[scale=.174]{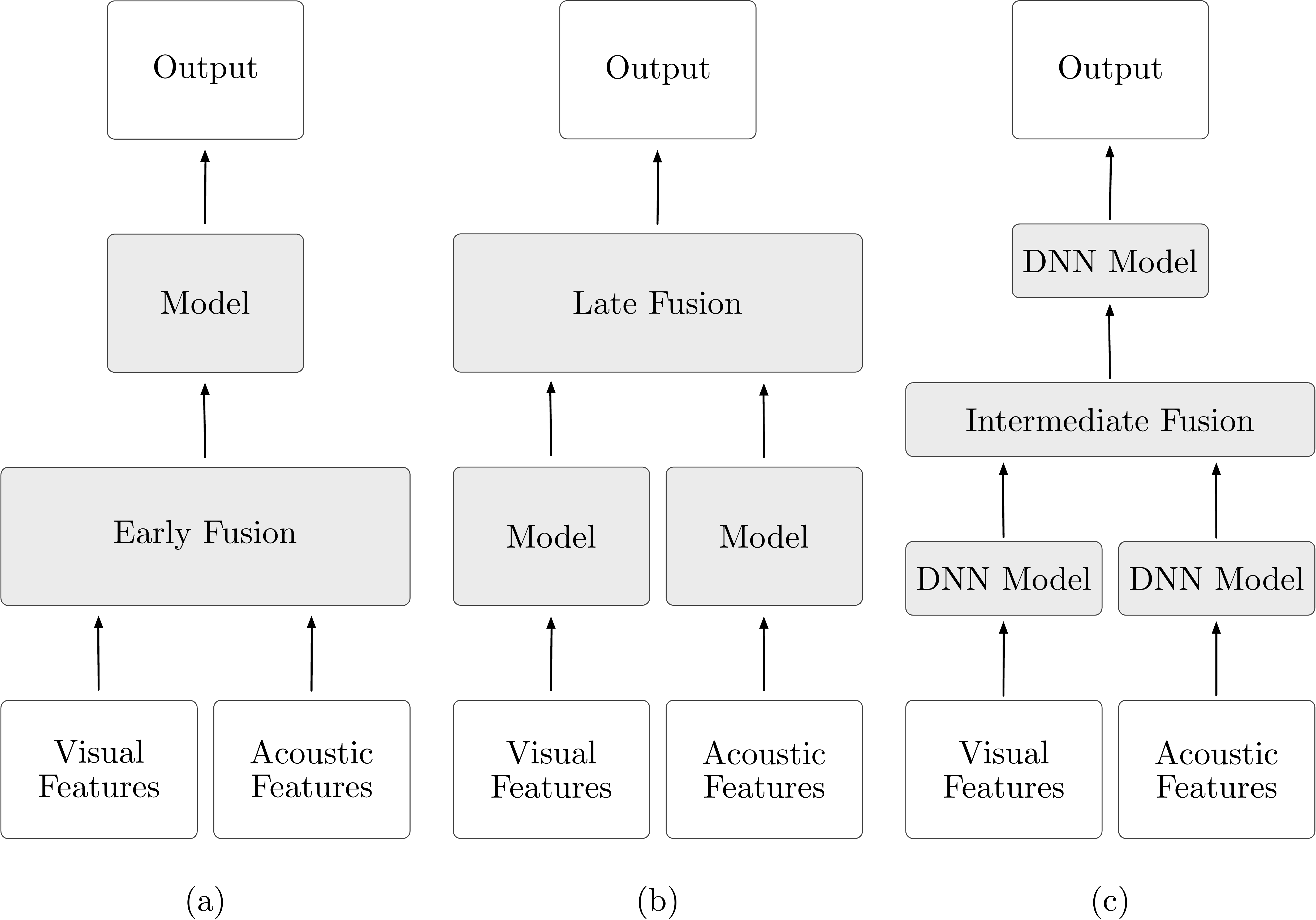}
	\caption{AV fusion paradigms. (a) Early fusion. (b) Late fusion. (c) Intermediate fusion. DNN model indicates a generic deep neural network model.}
	\label{fig:fusion_types}
\end{figure}

\subsection{{Audio-Visual Synchronisation}} 

{When acoustic and visual signals are recorded with different equipment, an AV synchronisation problem might occur. In other words, audio and video might not be temporally aligned. Humans can detect a lack of synchronisation when the audio leads the video by more than 45 ms or when the video is advanced with respect to the audio by more than 125 ms} \cite{itu1359_1}. {AV synchronisation has an impact also on AV-SE and AV-SS performance} \cite{afouras2018conversation}. {Since, in most existing works, the datasets used to train and evaluate AV-SE and AV-SS approaches are properly synchronised, the problem of temporal alignment for AV signals is usually not addressed. In fact, when the audio leads the video or vice versa, it is possible to pre-process the data using the approach proposed in} \cite{chung2016out}. {However, this method might fail at low SNRs} \cite{afouras2018conversation}. 

{Even when AV signals are temporally aligned, there might still be a need to synchronise acoustic and visual features because the two signals are sampled at different rates. As reported in Section} \ref{subsec:ac_f}, {most AV-SE and AV-SS systems use a TF representation of the acoustic signals. In this case, the audio frame rate is determined by the window size and the hop length chosen for the STFT and usually differs from the video frame rate. A common way to solve this problem is to upsample the video frames to match the temporal dimension of the acoustic features} \cite{afouras2019my}. {In this respect, the use of time-domain acoustic signals, as done in recent end-to-end deep-learning-based systems, might be beneficial, since it poses fewer constraints than the STFT.}

\subsection{{Traditional Fusion Paradigms}}

The traditional multimodal fusion approaches are generally grouped into two classes, based on the processing level at which the fusion occurs \cite{ramachandram2017deep, liu2018learn}: \emph{early fusion} and \emph{late fusion}. 

{As shown in} Figure \ref{fig:fusion_types}, early fusion {consists of combining} the information of the different modalities into a joint representation at the feature level. The main advantage is that the correlation between audio and video can be exploited with a single model at a very early stage, making the system more robust if compared to another one that processes the two modalities separately and combines them only at a later stage. Evidence in speech perception suggests that also in humans the AV integration occurs at a very early stage \cite{schwartz2002audio}. The disadvantage of early fusion is that usually the features of the two modalities are inherently different. Therefore, appropriate techniques for feature normalisation, transformation and synchronisation need to be developed.  

Late fusion, on the other hand, consists of combining the modalities  only at the decision level, after that the acoustic and visual information is processed separately with two different models (cf. Figure \ref{fig:fusion_types}). Although, from a theoretical perspective, early fusion would be preferable for the reasons mentioned above, late fusion is often used in practice for two reasons: it is possible to use unimodal models designed and validated over the years to achieve the best performance for each modality \cite{joze2019mmtm}; 
it is easy to perform late fusion, because the data processed from the two modalities belongs to the same domain, being different estimates of the same quantity.

\subsection{{Fusion Paradigms with Deep Learning}}

Although some AV-SE and AV-SS works showed that deep learning offers the possibility to perform both early \cite{morrone2019face} and late \cite{khan2018using, gabbay2018seeing} fusion, the majority of existing systems (e.g. \cite{gabbay2018visual, ephrat2018looking, afouras2018conversation, joze2019mmtm}) exploited the flexibility of deep learning techniques and fused the different unimodal representations into a single hidden layer. This fusion strategy is known as \emph{intermediate fusion}~\cite{ramachandram2017deep} (cf. Figure \ref{fig:fusion_types}).

\begin{table}
\caption{{List of fusion techniques in audio-visual speech enhancement and separation papers.}}
\centering
\resizebox{0.48\textwidth}{!}{%
\begin{tabular}{l l}
\toprule
Fusion Techniques & AV-SE/SS papers \\
\midrule
 Concatenation-based & \cite{aldeneh2020self, arriandiaga2019audio, afouras2018conversation, afouras2019my, adeel2019contextual, adeel2019novel, chung2020facefilter, chuang2020lite, ephrat2018looking, gogate2018dnn}\\
 & \cite{gogate2019cochleanet, gu2020multi, hou2017audio, hou2016audio, ideli2019visually, ideli2019multi, inan2019evaluating, 9054180, li2020visual}\\ 
&\cite{ lu2018listen, lu2019audio, luo2019audio, michelsanti2019training, michelsanti2019deep, michelsanti2019effects, morrone2019face}\\
&\cite{pasa2019joined, owens2018audio, qu2020multimodal, sadeghi2019audio, sadeghi2019robust, tan2019audio, sun2020attention} \\
&\cite{wu2016multi, wu2019time, xu2020neural} \\
 Addition-based \rule{0pt}{2.5ex} & \cite{afouras2019my, khan2018using}\\
 Product-based \rule{0pt}{2.5ex} & \cite{ochiai2019multimodal, 9053033, lu2018listen}\\
 Squeeze-excitation fusion \rule{0pt}{2.5ex} & \cite{joze2019mmtm, iuzzolino2020av}\\
 Attention-based  \rule{0pt}{2.5ex} & \cite{9054180, ochiai2019multimodal, chung2020facefilter, sun2020attention, gu2020multi} \\
 Integration within a Wiener \rule{0pt}{2.5ex} & \cite{adeeltowards, adeel2019lip, adeel2019contextual, adeel2019novel} \\
 filtering framework \\
\bottomrule
\end{tabular}}
\label{tab:avss_fusion_techniques}
\end{table}

\begin{figure}
	\centering
		\includegraphics[scale=.18]{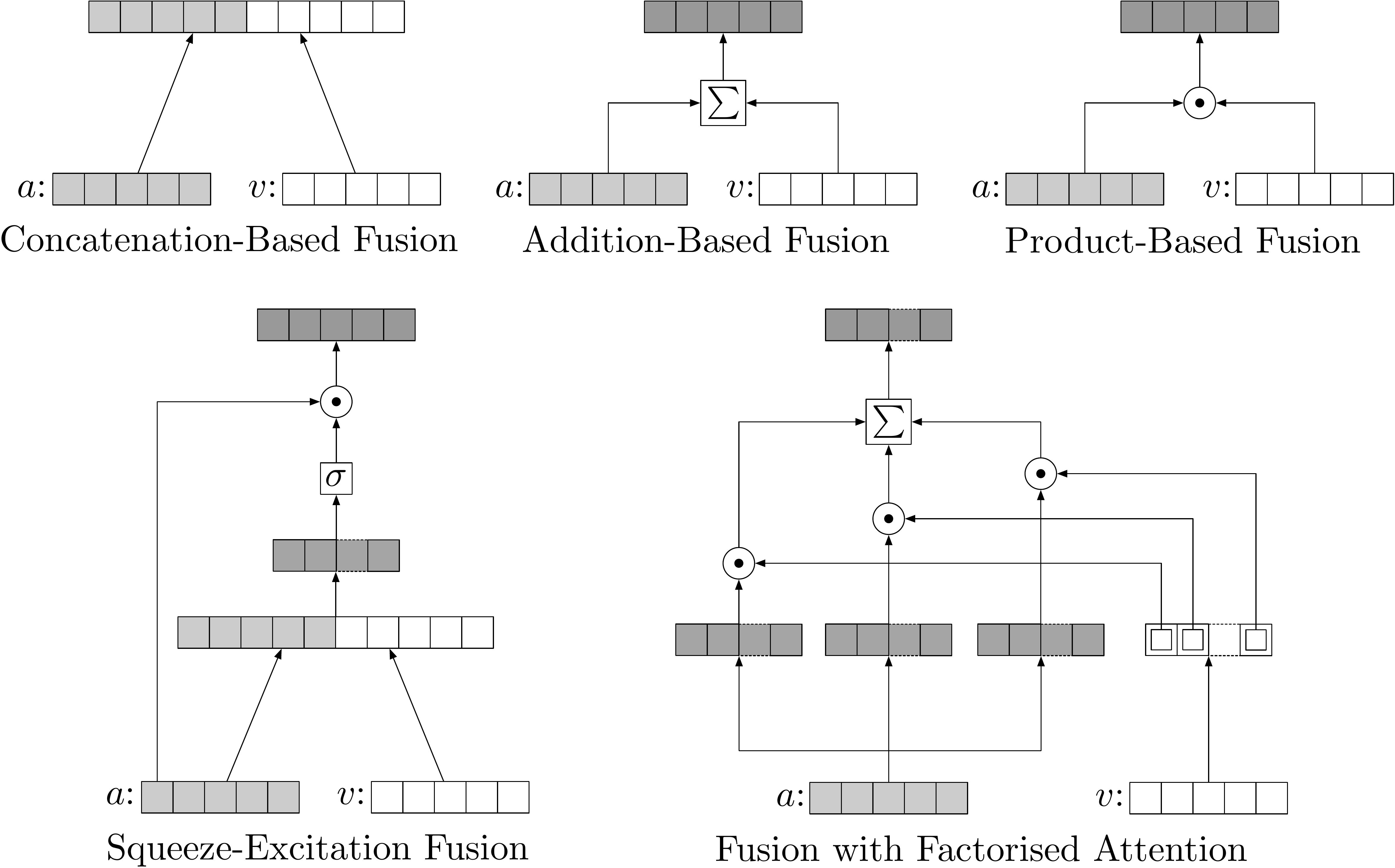}
	\caption{{Simplified graphical representation of the main fusion techniques used in audio-visual speech enhancement and separation systems. More details regarding the specific operations in squeeze-excitation fusion and fusion with factorised attention can be found in} \cite{iuzzolino2020av} and in \cite{gu2020multi}, {respectively. The symbols $a$ and $v$ indicate acoustic and visual feature vectors, respectively.}}
	\label{fig:fusion_techniques}
\end{figure}

Besides the level at which the AV integration occurs, it is important to consider the way in which this integration is performed. Table \ref{tab:avss_fusion_techniques} {reports a list of the fusion techniques used in the literature, and the most important ones are represented in Figure} \ref{fig:fusion_techniques}. The preferred way to fuse the information in AV-SE and AV-SS systems is through \emph{concatenation}. Although this approach is easy to implement, it comes with some potential problems. When two modalities are concatenated, the system uses them simultaneously and treats them in the same way. This means that although, in principle, a deep-learning-based system trained with a very large amount of data should be able to distinguish the cases in which the two modalities are complementary or in conflict \cite{liu2018learn}, in practice we often experience that one modality (not necessarily the most reliable in a given scenario) tends to dominate over the other \cite{feichtenhofer2016convolutional, gabbay2018visual}, causing a performance degradation. In AV-SE and AV-SS the acoustic {modality} is the one that dominates \cite{hou2017audio, gabbay2018visual}. This is something that might happen also for the approaches that employ an \emph{addition-based fusion}, in which the representations of the multimodal signals are added, with or without weights, not dealing explicitly with the aforementioned issues. Research has been conducted to investigate several possible methods to avoid that one modality dominates over the other. We provide some examples in the following.

Hou et al. \cite{hou2017audio} adopted two strategies. First, they forced the system under development to use both modalities by learning the target speech and the video frames of the speaker mouth at the same time. However, this approach alone does not guarantee that the network discovers AV correlations: it might happen that the network automatically learns to use some hidden nodes to process only the audio modality, and other nodes to process only the video modality. To avoid this selective behaviour, the second strategy adopted in \cite{hou2017audio} was a multi-style training approach \cite{ngiam2011multimodal, chung2017lip}, in which one of the input modalities could be randomly zeroed out. Gabbay et al. \cite{gabbay2018visual} introduced a new training procedure, which consisted of including training samples, in which the noise signal added to the target speech was, in fact, another utterance from the target speaker. Since it is hard to separate overlapping sentences from the same speaker using only the acoustic modality, the network learned to exploit the visual features better. Morrone et al. \cite{morrone2019face} proposed a two-stage training procedure: first, a network was forced to use visual information because it was trained to learn a mapping between the visual features and a target mask to be applied to the noisy spectrogram; then, a new network used the acoustic features together with the visually-enhanced spectrogram obtained from the previous stage to further enhance the speech signal. Wang et al. \cite{9053033} trained two networks separately for each modality to learn target masks and used a gating network to perform a \emph{product-based} fusion, keeping the system performance lower-bounded by the results of the AO network. This approach guaranteed good performance also at high SNRs, where many AV systems fail because acoustic information, which is very strong, and visual information, which is rather weak, is strongly coupled with early or intermediate fusion \cite{9053033}. Joze et al. \cite{joze2019mmtm} and Iuzzolino and Koishida \cite{iuzzolino2020av} proposed the use of squeeze-excitation blocks which generalised the work in \cite{hu2018squeeze} for multimodal applications. In particular, each block consisted of two units \cite{joze2019mmtm}: a squeeze unit that provided a joint representation of the features from each modality; an excitation unit which emphasised or suppressed the multimodal features from the joint representation based on their importance.

In order to softly select the more informative modality for AV-SE and AV-SS, \emph{attention-based fusion} mechanisms have also been investigated in several works \cite{9054180, ochiai2019multimodal, chung2020facefilter, sun2020attention, gu2020multi}. The attention mechanism \cite{bahdanau2015neural} was introduced in the field of natural language processing to improve sequence-to-sequence models \cite{cho2014learning, sutskever2014sequence} for neural machine translation. A sequence-to-sequence architecture consists of RNNs organised in an encoder, which reads an input sequence and compresses it into a context vector of a fixed length, and a decoder, which produces an output (i.e. the translated input sequence) considering the context vector generated by the encoder. Such a model fails when the input sequence is long, because the fixed-length context vector acts as a bottleneck. Therefore, Bahdanau et al. \cite{bahdanau2015neural} proposed to use a context vector that preserved the information of all the encoder hidden cells and allowed to align source and target sequences. In this case, the model could attend to salient parts of the input. Besides neural machine translation \cite{bahdanau2015neural, luong2015effective, vaswani2017attention}, attention was later successfully applied to various tasks, like image captioning \cite{vinyals2015show, xu2015show}, speech recognition \cite{chorowski2015attention} and speaker verification \cite{zhang2016end}. In the context of AV-SE and AV-SS, two representative works are \cite{chung2020facefilter} and \cite{gu2020multi}. In \cite{chung2020facefilter}, temporal attention \cite{lin2017structured} was used, motivated by the fact that different acoustic frames need different degrees of separation. For example, the frames where only the target speech is present should be treated differently from the frames containing overlapped speech or only the interfering speech. In \cite{gu2020multi}, a rule-based attention mechanism \cite{gu2019neural} was employed to take into account the fact that the significance of each information cue depended on the specific situation that the system needed to analyse. For example, when the speakers were close to each other, spatial and directional features did not provide high discriminability. Therefore, when the angle difference between the speakers was small, the attention weights allowed the model to selectively attend to the more salient cues, i.e. the spectral content of the audio and the lip movements. In addition, a factorised attention was adopted to fuse spatial information, speaker characteristics and lip information at embedding level. The model first factorised the acoustic embeddings into a set of subspaces (e.g., phone and speaker subspaces) and then used information from other cues to fuse them with selective attention.



\begin{table}
\caption{{List of training targets and objective functions in audio-visual speech enhancement and separation papers.}}
\centering
\resizebox{0.48\textwidth}{!}{%
\begin{tabular}{l l l l}
\toprule
Training Targets & AV-SE/SS papers \\
\midrule
 Magnitude spectrogram (DM) &  \multicolumn{3}{l}{\cite{adeeltowards, adeel2019lip, adeel2019contextual, adeel2019novel, chuang2020lite, michelsanti2019training, pasa2019joined, owens2018audio, hou2016audio, hou2017audio, tan2019audio, gabbay2018visual}}\\
&\multicolumn{3}{l}{ \cite{wu2016multi}}\\
 Phase \rule{0pt}{2.5ex} & \multicolumn{3}{l}{\cite{9054180, owens2018audio, afouras2018conversation, afouras2019my}} \\
 Mask:  \rule{0pt}{2.5ex} & MA: & IM: & Other: \\
 \quad- IBM \rule{0pt}{2.5ex} &   \quad \cite{gogate2018dnn, gogate2019cochleanet} & \quad $\,$ -- & \quad \cite{khan2018using, gabbay2018seeing}  \\
 &&&\quad \cite{lu2018listen, lu2019audio} \\
 \quad- TBM \rule{0pt}{2.5ex} &   \quad \cite{morrone2019face} &  \quad $\,$ -- & \quad \cite{gabbay2018seeing} \\ 
 \quad- PBM  \rule{0pt}{2.5ex} & \quad\cite{9053033} & \quad $\,$ -- & \quad $\,$ --  \\
 \quad- IRM  \rule{0pt}{2.5ex} &  \quad\cite{aldeneh2020self, 9053033} & \quad $\,$ -- & \quad \cite{khan2018using, gabbay2018seeing} \\ 
 \quad- IAM  \rule{0pt}{2.5ex} &  \quad\cite{michelsanti2019training, michelsanti2019effects} & \quad \cite{ afouras2018conversation, afouras2019my, arriandiaga2019audio, chung2020facefilter} & \quad $\,$ -- \\
 &\quad\cite{michelsanti2019deep}&\quad\cite{ephrat2018looking, iuzzolino2020av, joze2019mmtm} \\
 &&\quad\cite{9054180, michelsanti2019training, morrone2019face} \\
&&\quad \cite{ochiai2019multimodal,qu2020multimodal}\\
 \quad- Ratio mask \rule{0pt}{2.5ex} & \quad $\,$ -- & \quad $\,$ -- & \quad\cite{gu2020multi, tan2019audio} \\
&&&\quad \cite{xu2020neural}\\
 \quad- PSM  \rule{0pt}{2.5ex} & \quad\cite{michelsanti2019training} & \quad \cite{li2020visual, michelsanti2019training, ideli2019multi} & \quad $\,$ -- \\
 \quad- CRM  \rule{0pt}{2.5ex} & \quad\cite{luo2019audio} & \quad \cite{ephrat2018looking, ideli2019multi, inan2019evaluating} & \quad \cite{xu2020neural} \\
 && \quad \cite{sun2020attention}\\
 Waveform \rule{0pt}{2.5ex} & \multicolumn{3}{l}{\cite{ideli2019visually, wu2019time}} \\
 Mouth frames \rule{0pt}{2.5ex} & \multicolumn{3}{l}{\cite{hou2017audio}} \\
 Compressed mouth frames \rule{0pt}{2.5ex} & \multicolumn{3}{l}{\cite{chuang2020lite}} \\
\toprule
Objective Functions & AV-SE/SS papers \\
\midrule
 MSE & \multicolumn{3}{l}{\cite{adeeltowards, adeel2019lip, adeel2019contextual, adeel2019novel, aldeneh2020self, arriandiaga2019audio, chuang2020lite, chung2020facefilter, gabbay2018visual, hou2016audio, hou2017audio, ideli2019multi, inan2019evaluating}}\\
&\multicolumn{3}{l}{\cite{ khan2018using, li2020visual, michelsanti2019training, michelsanti2019effects, michelsanti2019deep, morrone2019face, pasa2019joined, ochiai2019multimodal, luo2019audio}}\\
&\multicolumn{3}{l}{\cite{  9053033, tan2019audio, sun2020attention, wu2016multi}} \\
 MAE \rule{0pt}{2.5ex} & \multicolumn{3}{l}{\cite{iuzzolino2020av, aldeneh2020self, pandey2018adversarial, owens2018audio, afouras2018conversation, afouras2019my, joze2019mmtm}} \\
 Cosine distance/similarity \rule{0pt}{2.5ex} & \multicolumn{3}{l}{\cite{aldeneh2020self, 9054180, afouras2018conversation, afouras2019my}} \\ 
 Cross entropy \rule{0pt}{2.5ex} & \multicolumn{3}{l}{\cite{ideli2019multi, 9053033, morrone2019face, gogate2019cochleanet, gogate2018dnn}}\\
 SI-SDR$^a$ \rule{0pt}{2.5ex} & \multicolumn{3}{l}{\cite{gu2020multi, ideli2019visually, wu2019time, tan2019audio, xu2020neural}} \\
 Multitask learning \rule{0pt}{2.5ex} & \multicolumn{3}{l}{\cite{9053033, chung2020facefilter, ochiai2019multimodal, pasa2019joined, hou2017audio}} \\ 
 CTC loss \rule{0pt}{2.5ex} & \multicolumn{3}{l}{\cite{pasa2019joined}} \\
 Speaker representation loss \rule{0pt}{2.5ex} & \multicolumn{3}{l}{\cite{chung2020facefilter}} \\
 PIT \rule{0pt}{2.5ex} & \multicolumn{3}{l}{\cite{ideli2019multi, owens2018audio}}\\
 Deep clustering \rule{0pt}{2.5ex} & \multicolumn{3}{l}{\cite{lu2018listen, lu2019audio}}\\
 Triplet loss \rule{0pt}{2.5ex} & \multicolumn{3}{l}{\cite{lu2018listen}} \\  
\bottomrule
  \multicolumn{3}{l}{$^a$Applied to the time-domain signal.}\\
\end{tabular}}
\label{tab:avss_tt_of}
\end{table}

For completeness, it is relevant to mention approaches that tried to leverage both deep-learning-based and knowledge-based models. For example, Adeel et al. \cite{adeel2019lip} used a deep-learning-based model to learn a mapping between the video frames of the target speaker and the filterbank audio features of the clean speech. The estimated speech features were subsequently used in a Wiener filtering framework to get enhanced short-time magnitude spectra of the speech of interest. This approach was extended in \cite{adeel2019contextual}, where both acoustic and visual modalities were used to estimate the filterbank audio features of the clean speech to be employed by the Wiener filter. The combination of deep-learning-based and knowledge-based approaches was leveraged not only in a single-microphone setup, but also for multi-microphone AV-SS. In \cite{xu2020neural}, a jointly trained combination of a deep learning model and a beamforming module was used. Specifically, a multi-tap minimum variance distortionless response (MVDR) was proposed with the goal of reducing the nonlinear speech distortions that are avoided with a MVDR beamformer \cite{capon1969high}, but inevitable for pure neural-network-based methods. With the jointly trained multi-tap MVDR, significant improvements of ASR accuracy could be achieved compared to the pure neural-network-based methods for the AV-SS task.


\subsection{{Shortcomings and Future Research}}

The fusion strategies and the design of neural network architectures experimented by researchers still require a lot of expertise. This means that, despite the number of works on AV-SE and AV-SS, researchers might not have explored the best architectures for data fusion. A way to deal with this issue is to investigate the possibility for a more general learning paradigm that focuses not only on determining the parameters of a model, but also on automatically exploring the space of the possible fusion architectures \cite{ramachandram2017deep}.

{In addition, future work should focus on techniques that take into account possible temporal misalignments of AV signals, which might make multimodality fusion critical.}



\section{{Training Targets and Objective Functions}} \label{subsec:tt_of}

As shown in Figure \ref{fig:avss_characteristics}, two other important elements of AV-SE and AV-SS systems are training targets, i.e. the desired outputs of deep-learning-based models, and objective functions, which provide a measure of the distance between the training targets and the actual outputs of the systems. Here, we discuss the adoption of the various training targets and objective functions for AV-SE and AV-SS comprehensively listed in Table \ref{tab:avss_tt_of}, using the taxonomy proposed in \cite{michelsanti2019training}.

\begin{figure}
	\centering
		\includegraphics[scale=.187]{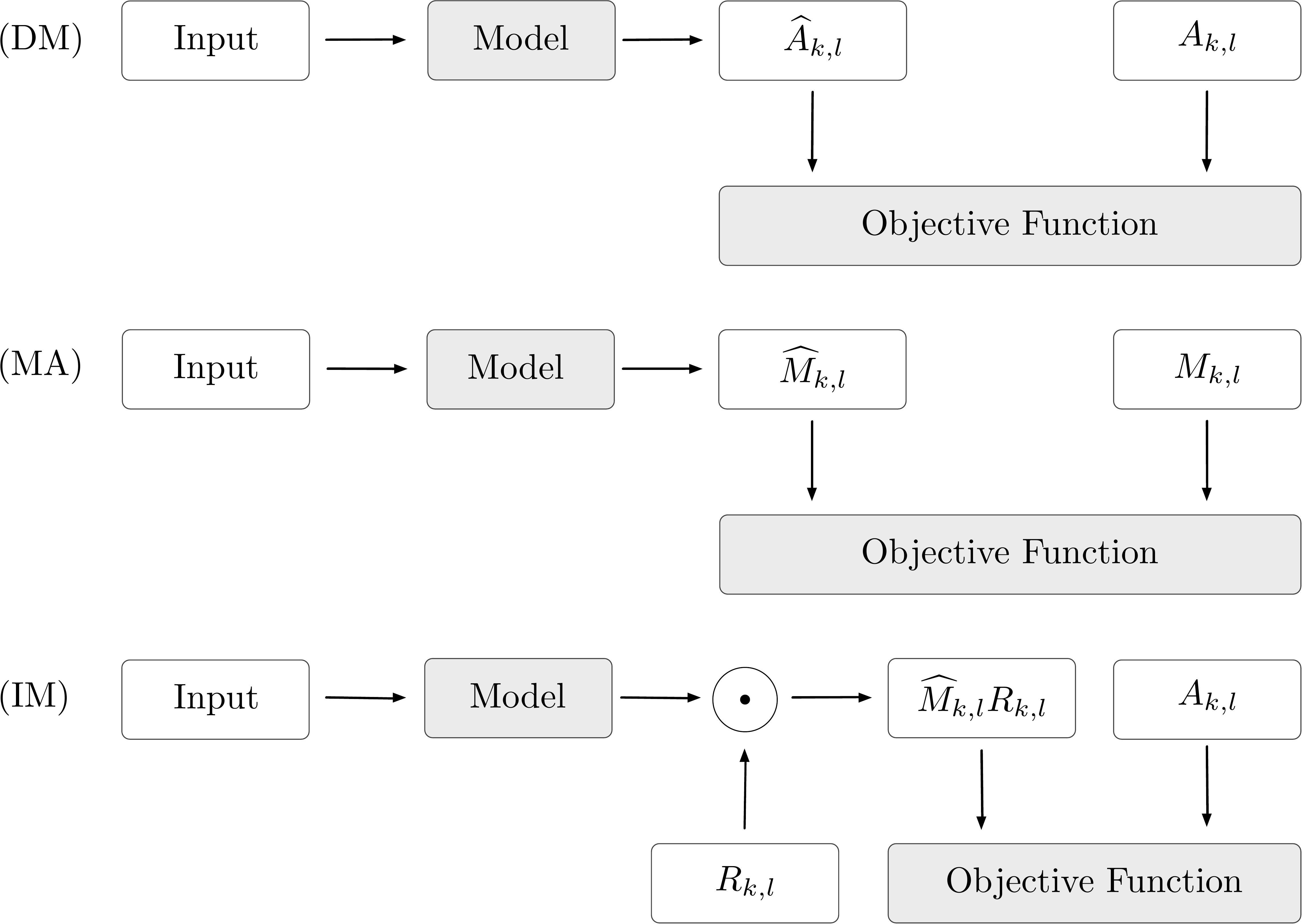}
	\caption{Illustration of direct mapping (DM), mask approximation (MA) and indirect mapping (IM) approaches. In the specific case of IM, the figure shows the estimation of the ideal amplitude mask. Similar illustrations can be made for different masks.}
	\label{fig:mapping}
\end{figure}

\subsection{{Direct Mapping}}

Following the terminology of Eq. (\ref{eq:sig_mod_2TF}) introduced in Section~\ref{subsec:prob_f} (the extension to SS is straightforward), let $A_{k,l} = \left| X(k,l) \right|$, $V_{k,l} = \left| D(k,l) \right|$ and $R_{k,l} = \left| Y(k,l) \right|$ indicate the magnitude of the STFT coefficients for the clean speech, the noise and the noisy speech signals, respectively. A common way to perform the enhancement is by \emph{direct mapping} (DM) \cite{sun2017multiple} (cf. Figure \ref{fig:mapping}): a system is trained to minimise an objective function reflecting the difference between the output, $\widehat{A}_{k,l}$, and the ground truth, $A_{k,l}$. The most frequently used objective function is the \emph{mean squared error} (MSE), whose minimisation is equivalent to maximising the likelihood of the data under the assumption of normal distribution of the errors. 
Alternatively, some AV models, such as \cite{owens2018audio}, have been trained with the \emph{mean absolute error} (MAE), experimentally proved to increase the spectral detail of the estimates and obtain higher performance if compared to MSE \cite{pandey2018adversarial, michelsanti2017conditional}. 

 In order to reconstruct the time-domain signal, an estimate of the target short-time phase is also needed. The noisy phase is usually combined with $\widehat{A}_{k,l}$, since it is the optimal estimator of the target short-time phase \cite{ephraim1984speech}, under the assumption of Gaussian distribution of speech and noise. However, choosing the noisy phase for speech reconstruction poses limitations to the achievable performance of a system. Iuzzolino and Koishida \cite{iuzzolino2020av} reported a significant improvement in terms of PESQ and STOI when their system used the target phase instead of the noisy phase to reconstruct the signal. This suggests that modelling the phase could be important in AV applications and some research \cite{9054180, owens2018audio, afouras2018conversation, afouras2019my} has moved towards this direction. Specifically, Owens and Efros \cite{owens2018audio} predicted both the target magnitude log spectrogram and the target phase with their model. Afouras et al. \cite{afouras2018conversation} designed a sub-network to specifically predict a residual which, when added to the noisy phase, allowed to estimate the target phase. In this case, the phase sub-network was trained to maximise the \emph{cosine similarity} between the prediction and the target phase, in order to take into account the angle between the two. The experiments showed that using the phase estimate was better than using the phase of the input mixture, although there was still room for improvements to match the performance obtained with the ground truth phase.

\subsection{{Mask Approximation}}

An alternative approach to DM consists of using a deep-learning-based model to get an estimate $\widehat{M}_{k,l}$ of a \emph{mask}, $M_{k,l}$. To reconstruct the clean speech signal during inference, $\widehat{M}_{k,l}$  needs to be element-wise multiplied with a TF representation of the noisy signal \cite{wang2014training, michelsanti2019training}. This approach is known as \emph{mask approximation} (MA), and an illustration of it is shown in Figure \ref{fig:mapping}.

In the literature, several masks have been defined in the context of AO-SE \cite{wang2014training, wang2018supervised} and then adopted for AV-SE and AV-SS. One way to build a TF mask is by setting its TF units to binary values according to some criterion. An example is the \emph{ideal binary mask} (IBM) \cite{wang2014training}, defined as:
\begin{equation}
M^{IBM}_{k,l} = \begin{cases} 1 & \frac{ A_{k,l} }{V_k,l } > \Gamma(k) \\[3pt]
0 & \text{ otherwise }\end{cases}
\label{eq:ibm}
\end{equation}
\noindent where $\Gamma(k)$ indicates a predefined threshold. Later, other binary masks have been defined, such as the \emph{target binary mask} (TBM) \cite{kjems2009role, wang2014training} and the \emph{power binary mask} (PBM) \cite{9053033}. They have all been adopted as training targets in AV approaches \cite{gogate2018dnn, gogate2019cochleanet, morrone2019face, 9053033} using the \emph{cross entropy loss} as objective function. 

Besides binary masks, which are based on the principle of classifying each TF unit of a spectrogram as speech or noise dominated, continuous masks have been introduced for soft decisions. An example is the \emph{ideal ratio mask} (IRM) \cite{wang2014training}:
\begin{equation}
M^{IRM}_{k,l} = \left( \frac{A^2_{k,l}}{A^2_{k,l}+V^2_{k,l}} \right)^\beta,
\label{eq:irm}
\end{equation}
\noindent where $\beta$ is a scaling parameter. It is worth mentioning that this mask is heuristically motivated, although its form for $\beta=1$ has some resemblance with the Wiener filter \cite{loizou2013speech, wang2018supervised}. IRM has been adopted as training target for a few AV models \cite{9053033, aldeneh2020self}, using either MSE or MAE as objective function. Aldeneh et al. \cite{aldeneh2020self} proposed the use of a hybrid loss which combined MAE and cosine distance to overcome the limitations of MSE, getting sharp results and bypassing the assumption of statistical independence of the IRM components that the use of MSE or MAE alone would imply. 

The IRM does not allow to perfectly recover the magnitude spectrogram of the target speech signal when multiplied with the noisy spectrogram. Hence, the \emph{ideal amplitude mask} (IAM) \cite{wang2014training} was introduced:
\begin{equation}
M^{IAM}_{k,l} = \frac{A_{k,l}}{R_{k,l}}.
\label{eq:iam}
\end{equation}
As we discussed previously, the noisy phase is often used to reconstruct the time-domain speech signal. All the masks that we mentioned above do not take the phase mismatch between noisy and target signals into account. Therefore, the \emph{phase sensitive mask} (PSM) \cite{erdogan2015phase, wang2018supervised} and the \emph{complex ratio mask} (CRM) \cite{williamson2016complex, wang2018supervised} have been proposed. PSM is defined as:
\begin{equation}
M^{\text{PSM}}_{k,l}~=~\frac{A_{k,l}}{R_{k,l}}\cos(\theta_{k,l}),
\label{eq:psm}
\end{equation}
and tries to compensate for the phase mismatch by introducing a factor, $\cos(\theta_{k,l})$, which is the cosine of the phase difference between the noisy and the clean signals. CRM is the only mask that allows to perfectly reconstruct the complex spectrogram of the clean speech when applied to the complex noisy spectrogram, i.e.:
\begin{equation}
X(k,l)~=~M^{\text{CRM}}_{k,l} * Y(k,l),
\label{eq:crm}
\end{equation}
where $*$ denotes the complex multiplication, and $M^{\text{CRM}}_{k,l}$ indicates the CRM. IAM, PSM and CRM can be found in several AV systems \cite{michelsanti2019training, michelsanti2019effects, michelsanti2019deep, luo2019audio}, adopting MSE as objective function.

MA is usually preferred to DM. The reason is that a mask is easier to estimate with a neural network \cite{wang2014training, erdogan2017deep, michelsanti2019training}. {An exception is AV speech dereverberation (SD), which is addressed only by Tan et al.} \cite{tan2019audio}, {although reverberations have an impact on the signal at the receiver end (cf.~the signal model presented in Section} \ref{subsec:prob_f}). In this case, the use of a mask, specifically a ratio mask, is discouraged for the two reasons reported in \cite{tan2019audio}. First, as seen in Eq. (\ref{eq:sig_mod_1_2}), reverberation is a convolutive distortion and as such it does not justify the use of ratio masking, which assumes that target speech and interference are uncorrelated \cite{wang2018supervised}. In addition, if a system consists of a cascade of SS and SD modules, such as \cite{tan2019audio}, a ratio mask applied in the SD stage would not be able to easily reduce the artefacts often introduced by SS, because they are correlated with the target speech signal \cite{tan2019audio}.

\subsection{{Indirect Mapping}}

An attempt to exploit the advantages of DM and MA at the same time is done by \emph{indirect mapping} (IM) \cite{weninger2014discriminatively, sun2017multiple} (cf.~Figure~\ref{fig:mapping}). In IM, the model outputs a mask, as in MA, because it is easier to estimate than a spectrogram as mentioned above, but the objective function is defined in the signal domain, as in DM.  A comparison between DM, MA and IM for AV-SE was conducted in \cite{michelsanti2019training}. In contrast to what one might expect, the results showed that IM did not obtain the best performance among the three paradigms, as observed also in \cite{sun2017multiple, weninger2014discriminatively} for AO systems. Weninger et al. \cite{weninger2014discriminatively} experimentally showed for AO-SS that IM alone performed worse than MA, but it was beneficial when used to fine-tune a system previously trained with the MA objective. Despite these results, AV-SE and AV-SS systems were often trained from scratch with the IM paradigm (cf. Table~\ref{tab:avss_tt_of}) obtaining good results. The reason is probably the use of large-scale datasets, which allowed an optimal convergence of the models.

\subsection{{Other Paradigms for Training Targets Estimation}}

Researchers experimented also with other ways than DM, MA and IM to estimate training targets with a neural network model. For example, Gabbay et al. \cite{gabbay2018seeing} and Khan et al. \cite{khan2018using} used an estimate of the clean magnitude spectrogram, obtained from visual features with a deep-learning-based model, to build a binary mask that could be applied to the noisy spectrogram. The approaches in \cite{tan2019audio, gu2020multi} can be considered an extension of IM for a time-domain objective. Specifically, a system was trained to output a TF ratio mask using SI-SDR as objective function applied to the reconstructed time-domain signals. The ratio mask obtained with this approach was different from the IAM, because it was not necessarily the one that allowed a perfect reconstruction of the clean magnitude spectrogram. In \cite{tan2019audio}, the system was also trained with an objective that combined MSE on the magnitude spectrograms  and SI-SDR on the waveform signals. An objective function in the time domain was also used in \cite{ideli2019visually, wu2019time}. In these cases, a system, inspired by {Conv-TasNet} \cite{luo2019conv}, was used to directly estimate the waveform of the target speech signal with the SI-SDR training objective.

\subsection{{Multitask Learning}}

Other AV systems \cite{pasa2019joined, chung2020facefilter, ochiai2019multimodal, hou2017audio} tried to improve SE and SS performance with \emph{multitask learning} (MTL) \cite{caruana1997multitask}, which consists of training a learning model to perform multiple related tasks. Pasa et al. \cite{pasa2019joined} investigated MTL using a joint system for AV-SE and ASR. They tried to either jointly minimise a SE objective, MSE, and an ASR objective, \emph{connectionist temporal classification} (CTC) \emph{loss} \cite{graves2006connectionist}, or alternate the training between an AV-SE {stage} and an ASR {stage}. The alternated training was reported to be the most effective. Chung et al. \cite{chung2020facefilter} used two objective functions to train their system: one was the MSE on magnitude spectrograms and the other was the \emph{speaker representation loss} \cite{mun2020sound} on embeddings from a network that extracted the speaker identity. Ochiai et al. \cite{ochiai2019multimodal} used a combination of losses that allowed their system to work even when either acoustic or visual cues of the speaker were not available. {Finally, Hou et al.} \cite{hou2017audio} {trained their system to both perform AV-SE and reconstruct visual features. As we explained before in Section} \ref{subsec:fusion}, {this approach forces the system to use visual information from the input.}

\subsection{{Source Permutation}}

A typical issue for SS is the so-called \emph{source permutation} \cite{hershey2016deep, yu2017permutation}. This problem occurs in speaker-independent SS systems and it is characterised by an inconsistent assignment over time of the separated speech signals to the sources. Two solutions have been proposed in AO settings: \emph{permutation invariant training} (PIT) \cite{yu2017permutation, kolbaek2017multitalker} and \emph{deep clustering} (DC) \cite{hershey2016deep, isik2016single, chen2017deep, luo2018speaker}. The idea behind PIT is to calculate the objective function for all the possible permutations of the sources and use the permutation associated with the lowest error to update the model parameters. In DC, an embedding vector is learned for each TF unit of the mixture spectrogram and is used to perform clustering to learn an IBM for SS. An extension of DC is the \emph{deep attractor network} \cite{chen2017deep}, which creates attractor points in the embedding space learned from the TF representation of the signal and estimates a soft mask from the similarity between the attractor points and the TF embeddings. Although some AV-SS systems used PIT or DC (cf. Table \ref{tab:avss_tt_of}), source permutation is less of a problem in {deep-learning-based} AV-SS, assuming that the target speakers are visible while they talk: visual information is a strong guidance for the systems and allows to automatically assign the separated speech signals to the correct sources.

\subsection{{Shortcomings and Future Research}}

Although many training targets and objective functions have already been investigated for AV-SE and AV-SS, we expect further improvements following several research directions, such as: the use of perceptually motivated objective functions; the estimation of binaural cues to preserve the spatial dimension also at the receiver end; a greater effort for {designing and estimating} time-domain training targets to perform end-to-end training.

\section{Related Research} \label{sec:related}

In this section, we consider two problems, \emph{speech reconstruction from silent videos} and \emph{audio-visual sound source separation for non-speech signals}, because the first is a special case of AV-SE in which the acoustic input is missing, while the second is {the complementary task of AV-SS within the AV sound source separation problem space, considering that} the {target signals are not speech signals}, {but, for example, sounds from musical instruments}\footnote{{Sometimes, the target signals include singing voices, which typically have different characteristics from speech.}}. 


{Models used for speech reconstruction from silent videos} can easily be adopted to estimate a mask for AV-SE and AV-SS. {An example is the one presented} in \cite{gabbay2018seeing}, {where TF masks, obtained by thresholding the reconstructed speech spectrograms, are used to filter the noisy spectrogram}. On the other hand, sound source separation {for non-speech signals} techniques can be adopted also for speech signals by re-training the deep-learning-based models on an AV speech dataset. In some cases, these techniques are domain-specific, such as \cite{gan2020music}, making the adoption to the speech domain hard. Nevertheless, the ways in which multimodal data is processed and fused can be of inspiration also for AV-SE and AV-SS.

\subsection{Speech Reconstruction from Silent Videos}

In some circumstances, the only reliable and accessible modality to understand the speech of interest is the visual one. Real-world scenarios of this kind include, for example: conversations in acoustically demanding situations like the ones occurring during a concert, where the sound from the loudspeakers tends to dominate over the target speech; teleconferences, in which sound segments are missing, e.g. due to audio packet loss \cite{morrone2020audio}; surveillance videos, generally recorded in a situation where the target speaker is acoustically shielded (e.g. with a window) from camera(s) and microphone(s). All these scenarios might be considered as an extreme case of AV-SE where the goal is to estimate the speech of interest from the silent video of a talking face.

In the literature, the problem of estimating speech from visual information is known as speech reconstruction from silent videos. {This task is hard because the information that can be captured by a frontal or a side camera is incomplete and cannot include: the excitation signal, i.e. the airflow signal immediately after the vocal chords; most of the the tongue movements. In particular, not having access to tongue movements is critical for speech synthesis, because they are very important for the generation of several speech sounds. For this reason, attempts were made to exploit silent articulations not visible with regular cameras using other sensors}\footnote{The data used in these works include (but are not limited to) recordings obtained with electromagnetic articulography, electropalatography and laryngography sensors.}  \cite{kello2004neural, denby2010silent, jorgensen2010speech, hueber2016statistical}. {In this subsection we will not consider such silent speech interfaces because they greatly differ from the main topic of this overview. Instead, we focus on deep-learning-based approaches, as listed in Table} \ref{tab:speech_rec}, {that directly perform a mapping from videos captured with regular cameras to speech signals}.



\begin{table}
\caption{{Chronological list of} deep-learning-based approaches for speech reconstruction from silent videos. MV: multi-view. SI:~speaker-independent. VSR: visual speech recognition.}
\centering
\resizebox{0.48\textwidth}{!}{%
\begin{tabular}{l l l l l c c c}
\toprule
Paper   & Year & Input & Output & Model Info & MV & SI & VSR\\
\midrule
\cite{cornu2015reconstructing} & 2015 & 2-D DCT / AAM & LPC or  & GMM / FFNN & \xmark & \xmark & \xmark  \\
&& mouth &mel-filterbank \\
&&&amplitudes\\
\cite{le2017generating} \rule{0pt}{2.5ex} & 2017 & AAM & Codebook entries & FFNN / RNN &\xmark & \xmark & \xmark \\
&& mouth &(mel-filterbank \\
&&& amplitudes) \\
\cite{ephrat2017vid2speech} \rule{0pt}{2.5ex} & 2017 & Raw pixels  & LSP of LPC  & CNN, FFNN  &\xmark & \xmark & \xmark \\
&& face\\
\cite{ephrat2017improved} \rule{0pt}{2.5ex}& 2017 & Raw pixels, & Mel-scale and&CNN, FFNN, &\xmark&\xmark & \xmark \\
&& optical flow&linear-scale& BiGRU \\
&& face & spectrograms\\
\cite{akbari2018lip2audspec} \rule{0pt}{2.5ex}& 2018 & Raw pixels & AE features, & CNN, LSTM,  &\xmark&\xmark & \xmark \\
&& face & spectrogram & FFNN, AE\\
\cite{kumar2018harnessing} \rule{0pt}{2.5ex} & 2018 & Raw pixels & LSP of LPC & CNN, LSTM, & \cmark & \xmark & \xmark \\
&& mouth && FFNN \\
\cite{kumar2018mylipper} \rule{0pt}{2.5ex} & 2018 & Raw pixels & LSP of LPC &CNN, BiGRU,& \cmark & \xmark & \xmark \\
&& mouth &&  FFNN \\
\cite{kumar2019lipper} \rule{0pt}{2.5ex} & 2019 & Raw pixels & LSP of LPC & CNN, BiGRU,& \cmark & \cmark & \cmark \\
&& mouth && FFNN \\
\cite{takashima2019examplar} \rule{0pt}{2.5ex}& 2019 & Raw pixels & WORLD & CNN, FFNN  &\xmark&\xmark&\xmark \\
&& mouth & spectrum & \\
\cite{Vougioukas2019} \rule{0pt}{2.5ex}& 2019 & Raw pixels & Raw waveform & GAN, CNN,  &\xmark&\cmark&\xmark \\
&& mouth & & GRU\\
\cite{Uttam2019} \rule{0pt}{2.5ex} & 2019 & Raw pixels &AE features,& CNN, LSTM& \cmark & \cmark & \xmark \\
&& mouth & spectrogram & FFNN, AE\\
\cite{michelsanti2020vocoder} \rule{0pt}{2.5ex} & 2020 & Raw pixels  & WORLD & CNN, GRU, &\xmark&\cmark&\cmark \\
&& mouth / face & features &  FFNN\\
\cite{prajwal2020learning} \rule{0pt}{2.5ex} & 2020 & Raw pixels & mel-scale & CNN, LSTM &\cmark&\xmark&\xmark \\
&& face & spectrogram\\
\bottomrule
\end{tabular}}
\label{tab:speech_rec}
\end{table}


Le Cornu and Milner \cite{cornu2015reconstructing} were the first to employ a neural network to estimate a speech signal using only the silent video of a speaker's frontal face. They decided to base their system on STRAIGHT \cite{kawahara1999restructuring}, a vocoder which allows to perform speech synthesis from {a set of} time-varying parameters describing fundamental aspects of a given speech signal: fundamental frequency (F0), aperiodicity (AP) and spectral envelope (SP). Supported by the results of some previous works \cite{yehia1998quantitative, barker1999evidence, almajai2006analysis}, they assumed that only SP could be inferred from visual features. Therefore, AP and F0  were not estimated from the silent video, but artificially produced without taking the visual information into account, while SP was estimated with a Gaussian mixture model (GMM) and FFNN within a regression-based framework. As input to the models, two different visual features were considered, 2-D DCT and AAM, while the explored SP representations were linear predictive coding (LPC) coefficients and mel-filterbank amplitudes. While the choice of visual features did not have a big impact on the results, the use of mel-filterbank amplitudes allowed to outperform the systems based on LPC coefficients.

This work was extended in \cite{le2017generating}, where two improvements were proposed. First, instead of adopting a regression framework, visual features were used to predict a class label, which in turn was used to estimate audio features from a codebook. Secondly, the influence of temporal information was explored from a feature-level point of view, by grouping multiple frames, and from a model-level point of view, by using RNNs. The obtained improvement in terms of intelligibility was substantial, but the speech quality was still low, mainly because the excitation parameters, i.e. F0 and AP, were produced without exploiting visual cues.

Ephrat and Peleg \cite{ephrat2017vid2speech} moved away from a classification-based method as the one presented in \cite{le2017generating} and went back to a regression-based framework. Their approach consisted of predicting a line spectrum pairs (LSP) representation of LPC coefficients directly from raw visual data with a CNN, followed by two fully connected layers. Their findings demonstrated that: no hand-crafted visual features were needed to reconstruct the speaker's voice; using the whole face instead of the mouth area as input improved the performance of the system; a regression-based method was effective in reconstructing out-of-vocabulary words. Although the results were promising in terms of intelligibility, the signals sounded unnatural because Gaussian white noise was used as excitation to reconstruct the waveform from LPC features. Therefore, a subsequent study \cite{ephrat2017improved} focused on speech quality improvements. In particular, the proposed system was designed to get a linear-scale spectrogram from a learned mel-scale one with the post-processing network in \cite{wang2017tacotron}. The time-domain signal was then reconstructed combining an example-based technique similar to \cite{owens2016visually} with the Griffin-Lim algorithm \cite{griffin1984signal}. Furthermore, a marginal performance improvement was obtained by providing not only raw video frames as input, but also optical flow fields computed from the visual feed.

Another system was developed by Akbari et. al. \cite{akbari2018lip2audspec}, who tried to reconstruct natural sounding speech by learning a mapping between the speaker's face and speech-related features extracted by a pre-trained deep AE. The approach was effective and outperformed the method in \cite{ephrat2017vid2speech} in terms of speech quality and intelligibility.

The main limitation of these techniques was that they were employed to reconstruct speech of talkers observed by the model at training time. The first step towards a system that could generate speech from various speakers was taken by Takashima et al. \cite{takashima2019examplar}. They proposed an exemplar-based approach, where a CNN was trained to learn a high-level acoustic representation from visual frames. This representation was used to estimate the target spectrogram with the help of an audio dictionary. The approach could generate a different voice without re-training the neural network model, but by simply changing the dictionary with that of another speaker.

Prajwal et al. \cite{prajwal2020learning} developed a sequence-to-sequence system adapted from Tacotron 2 \cite{shen2018natural}. Although their goal was to learn speech patterns of a specific speaker from videos recorded on unconstrained settings, obtaining state-of-the-art performance, they also proposed a multi-speaker approach. In particular, they conditioned their system on speaker embeddings extracted from a reference speech signal as in \cite{jia2018transfer}. Although they could synthesise speech of different speakers, prior information was needed to get speaker embeddings. Therefore, this method cannot be considered a speaker-independent approach, but a speaker-adaptive one.

The challenge of building a speaker-independent system was addressed by Vougioukas et al. \cite{Vougioukas2019}, who developed a generative adversarial network (GAN) that could directly estimate time-domain speech signals from the video frames of the talker's mouth region. Although this approach was capable of reconstructing intelligible speech also in a speaker independent scenario, the speech quality estimated with PESQ was lower than that in \cite{akbari2018lip2audspec}. The generated speech signals were characterised by a low-power hum, presumably because the model output was a raw waveform, for which suitable loss functions are hard to find \cite{ephrat2017vid2speech}.

The method proposed in \cite{michelsanti2020vocoder} intended to still be able to reconstruct speech in a speaker independent scenario, but also to avoid artefacts similar to the ones introduced by the model in \cite{Vougioukas2019}. Therefore, vocoder features were used as training target instead of raw waveforms. Differently from \cite{cornu2015reconstructing, le2017generating}, the system adopted WORLD vocoder \cite{morise2016world}, which was proved to achieve better performance than STRAIGHT \cite{morise2018sound}, and was trained to predict all the vocoder parameters, instead of SP only. In addition, it also provided a VSR module, useful for all those applications requiring captions. The results showed that a MTL approach, where VSR and speech reconstruction were combined, was beneficial for both the estimated quality and the estimated intelligibility of the generated speech signal.

Most of the systems described above assumed that the speaker constantly faced the camera. This is reasonable in some applications, e.g. teleconferences. Other situations may require a robustness to multiple views and face poses. Kumar et al. \cite{kumar2018harnessing} were the first to make experiments in this direction. Their model was designed to take as input multiple views of the talker's mouth and to estimate a LSP representation of LPC coefficients for the audio feed. The best results in terms of estimated speech quality were obtained when two different views were used as input. The work was extended in  \cite{kumar2018mylipper}, where results from extensive experiments with a model adopting several view combinations were reported. The best performance was achieved with the combination of three angles of view (0$^\circ$, 45$^\circ$ and 60$^\circ$).

The systems in \cite{kumar2018harnessing} and in \cite{kumar2018mylipper} were personalised, meaning that they were trained and deployed for a particular speaker. Multi-view speaker-independent approaches were proposed in \cite{kumar2019lipper} and \cite{Uttam2019}. In both cases, a classifier took as input the multi-view videos of a talker and determined the angles of view from a discrete set of lip poses. Then, a decision network chose the best view combination and the reconstruction model to generate the speech signal. The main difference between the two systems was the audio representation used. While Uttam et al. \cite{Uttam2019} decided to work with features extracted by a pre-trained deep AE, similarly to \cite{akbari2018lip2audspec}, the approach in \cite{kumar2019lipper} estimated a LSP representation of LPC coefficients. In addition, Kumar et al. \cite{kumar2019lipper} provided a VSR module, as in \cite{michelsanti2020vocoder}. However, this module was trained separately from the main system and was designed to provide only one among ten possible sentence transcriptions, making it database-dependent and not feasible for real-time applications.

Despite the research done in this area, several critical points need to be addressed before speech reconstruction from silent videos reaches the maturity required for a commercial deployment. All the approaches in the literature except \cite{prajwal2020learning} presented experiments conducted in controlled environments. Real-world situations pose many challenges that need to be taken into account, e.g. the variety of lighting conditions and occlusions. Furthermore, before a practical system can be employed for unseen speakers, performance needs to improve considerably. At the moment, the results for the speaker-independent case are unsatisfactory, probably due to the limited number of speakers used in the training phase.

\subsection{Audio-Visual Sound Source Separation for Non-Speech Signals}

\begin{table}
\caption{{Chronological list of} deep-learning-based approaches for audio-visual source separation for non-speech signals. L:~Localisation.}
\centering
\resizebox{0.48\textwidth}{!}{%
\begin{tabular}{l l l c}
\toprule
Paper   & Year & Key Idea & L \\
\midrule
\cite{gao2018learning} & 2018  & Guide source separation with audio frequency bases & \xmark   \\
& & learned with a framework that maps to visual objects. &    \\
\cite{zhao2018sound} \rule{0pt}{2.5ex} & 2018 & Separate audio sources into components that can be & \cmark \\
& & localised in the video frames. &    \\
\cite{rouditchenko2019self} \rule{0pt}{2.5ex} & 2019 & Perform independent image co-segmentation and & \cmark \\
& & sound source separation for not synchronised data. &    \\
\cite{gao20192} \rule{0pt}{2.5ex} & 2019 & Use predicted binaural audio to aid sound source & \cmark \\
& & separation. &    \\
\cite{parekh2019identify} \rule{0pt}{2.5ex} & 2019 & Use of a multiple instance learning paradigm for & \cmark \\
& & separation and localisation of weakly-labeled data. &    \\
\cite{zhao2019sound} \rule{0pt}{2.5ex} & 2019 & Incorporate temporal motion information and employ & \cmark \\
& & a curriculum learning scheme for training. &    \\
\cite{xu2019recursive} \rule{0pt}{2.5ex} & 2019 & Do not separate the sounds independently to avoid    & \cmark \\
& & that acoustic components from the original mixture &    \\
& & get lost. &    \\
\cite{gao2019co} \rule{0pt}{2.5ex} & 2019 & Devise a new paradigm to use videos with  multiple & \xmark \\
& & (correlated) sounds during training. &    \\
\cite{slizovskaia2020conditioned} \rule{0pt}{2.5ex} & 2020 & Explore conditioning techniques with video stream  & \xmark \\
& & and weak labels. \\
\cite{gan2020music} \rule{0pt}{2.5ex} & 2020 & Use keypoint-based structured visual representations & \xmark \\
& & to model human-object interactions. \\
\cite{zhu2020visually} \rule{0pt}{2.5ex} & 2020 & Refine the separated sounds with cascaded opponent & \cmark \\
& & filtering. \\
\cite{zhu2020separating} \rule{0pt}{2.5ex} & 2020 & Use an appearance attention module for separation.  & \cmark \\
\bottomrule
\end{tabular}}
\label{tab:nonspeech_sep}
\end{table}

Sound source separation might involve signals different from speech. Imagine, for example, the task of extracting the individual sounds coming from different music instruments playing together. Although the signal of interest is not speech in this case, the approaches developed in this area can provide useful insights also for AV-SE and AV-SS.

Several works addressed AV source separation for non-speech signals. Similarly to other fields, classical methods \cite{barzelay2007harmony, casanovas2010blind, parekh2017motion, pu2017audio, parekh2017guiding} were recently replaced by deep-learning-based approaches, {that we listed in} Table \ref{tab:nonspeech_sep}. The first two works that concurrently proposed deep processing stages for the task under analysis were \cite{gao2018learning} and \cite{zhao2018sound}.

In \cite{gao2018learning}, a novel neural network for multi-instance multi-label learning (MIML) was used to learn a mapping between audio frequency bases and visual object categories. Disentangled audio bases were used to guide a non-negative matrix factorisation (NMF) framework for source separation. The method was successfully employed for in-the-wild videos containing a broad set of object sounds, such as musical instruments, animals and vehicles. NMF was also adopted in a later work by Parekh et al. \cite{parekh2019identify}, where both audio frequency bases and their activations were used, leveraging temporal information. In contrast to  \cite{gao2018learning}, the system could also perform visual \emph{localisation}, which is the task of detecting the sound sources in the visual input.

In \cite{zhao2018sound}, audio and video information were jointly used by a deep system called PixelPlayer to simultaneously localise the sound sources in the visual frames and acoustically separate them. The results of this technique sparked a particular interest in the research community, causing the development of several methods aiming at improving it further.

First of all, Rouditchenko et al. \cite{rouditchenko2019self} extended the work in \cite{zhao2018sound} for unsynchronised audio and video data. Their approach consisted of a network that learned disentangled acoustic and visual representations to independently perform visual object co-segmentation and sound source separation.

Then, PixelPlayer only considered semantic features extracted from the video frames. Appearance information is important as highlighted in \cite{zhu2020separating}, where the separation was guided with a single image, but higher performance is expected to be achieved when also motion information is exploited. Zhao et al. \cite{zhao2019sound} proposed to combine trajectory and semantic features to condition a source separation network. The system was trained with a curriculum learning scheme, consisting of three consecutive stages characterised by increasing levels of difficulty. This approach showed its effectiveness even for separating sounds of the same kind of musical instruments, an achievement not possible in \cite{zhao2018sound}. However, the trajectory motion cues are not able to accurately model the interactions between a human and an object, e.g. a musical instrument. For this reason, Gan et al. \cite{gan2020music} proposed to use keypoint-based structured visual representations together with the visual semantic context. In this way, they were able to achieve state-of-the-art performance. Motion information was also used in the form of optical flow and dynamic image \cite{bilen2016dynamic} by Zhu and Rahtu \cite{zhu2020visually}. Their approach refined the separated sounds in multiple stages within a framework called cascaded opponent filter (COF). In addition, they could achieve accurate sound source localisation with a sound source location masking (SSLM) network, following the idea in \cite{hu2019visualization}.

Especially when dealing with musical instruments, having a priori knowledge of the presence or absence of a particular instrument in a recording, i.e. weak labels, might be advantageous. Slizovskaia et al. \cite{slizovskaia2020conditioned} studied the problem of source separation conditioned with additional information, which included not only visual cues but also weak labels. Their investigation covered, among other aspects, neural network architectures (either U-Net \cite{ronneberger2015u} or multi-head U-Net (MHU-Net) \cite{doire2019interleaved}), conditioning strategies  (either feature-wise linear modulation (FiLM) \cite{dumoulin2018feature} or multiplicative conditioning), places of conditioning (at the bottleneck, at all the encoder layers or at the final decoder layer), context vectors (static visual context vector, visual-motion context vector and binary indicator vector encoding the instruments in the mixture) and training targets (binary mask or ratio mask).

The audio signals used in these systems are generally monaural. Inspired by the fact that humans benefit from binaural cues \cite{hawley2004benefit}, Gao and Grauman \cite{gao20192} proposed a method to exploit visual information with the aim of converting monaural audio into binaural audio. This conversion allowed to expose acoustic spatial cues that turned out to be helpful for sound source separation.

Zhao et al. \cite{zhao2018sound} separated the sound sources  in the observed mixture assuming that they were independent. This assumption can generate two main issues. The first is that the sum of the separated sounds might be different from the actual mixture, i.e. some acoustic components of the actual mixture might not be found in any outputs of the separation system. Therefore, Xu et al. \cite{xu2019recursive} proposed a novel method called MinusPlus network. The idea was to have a two-stage system in which: a minus stage recursively identified the sound with the highest energy and removed it from the mixture; a plus stage refined the removed sounds. The recursive procedure based on sound energy allowed to automatically handle a variable number of sound sources and made the sounds with less energy emerge. The second issue is related to the fact that training is usually performed following a paradigm in which distinct AV clips are randomly mixed. However, sounds that appear in the same scene are usually correlated, e.g. two musical instruments playing the same song. The use of training materials consisting of independent videos might hinder a deep network from capturing such correlations. Hence, Gao and Grauman \cite{gao2019co} introduced a new training paradigm, called co-separation, in which an association between consistent sounds and visual objects across pairs of training videos was learned. Exploring this aspect further and possibly overcoming the supervision paradigm used in most of the works in the literature by using real-world recordings and not only synthetic mixtures for training is an interesting future research direction that can easily be adopted also for AV-SE and AV-SS.

\section{{Audio-Visual Speech Corpora}} \label{sec:data}

One of the key aspects that allowed the recent progress and adoption of deep learning techniques for a range of different tasks is the availability of large-scale datasets. Therefore, the choice of a database is critical and it is determined by the specific purpose of the research that needs to be conducted. With this Section, our goal is to provide a non-exhaustive overview of existing resources which will hopefully help the reader to choose AV datasets that suit their purpose. In Table \ref{tab:databases}, we provide information regarding AV speech datasets, such as: year of publication, number of speakers, linguistic content, video resolution and frame rate, audio sample frequency, additional characteristics (e.g. recording settings) and the AV-SE and AV-SS papers in which the datasets are used for the experiments.

{Most of the datasets provide clean signals of several speakers. Researchers use these signals to create synthetic scenes with overlapping speakers and/or acoustic background noise from the AV speech databases or other sources. We expect that effort will be put in providing AV datasets containing speaker mixtures combined with real noise signals to have a benchmark for AV-SS in noise, like in AO-SS} \cite{wichern2019wham}.

\subsection{{Data in Controlled Environment}}
From Table \ref{tab:databases}, we notice that the two most commonly used databases in the area of deep-learning-based AV-SE and AV-SS are GRID \cite{cooke2006audio} and TCD-TIMIT \cite{harte2015tcd}. GRID consists of audio and video recordings  where 34 speakers (18 males and 16 females) pronounce 1000 sentences each. The data was collected in a \emph{controlled environment}: the speakers were placed in front of a plain blue wall inside an acoustically isolated booth and their face was uniformly illuminated. A GRID sentence has the following structure:  $<$command(4)$>$ $<$color(4)$>$ $<$preposition(4)$>$ $<$letter(25)$>$ $<$digit(10)$>$ $<$adverb(4)$>$, where the number of choices for each word is indicated in parentheses. Although the number of possible command combinations using such a sentence structure is high, the vocabulary is small, with only 51 words. This may pose limitations to the generalisation performance of a deep learning model trained with this database. Similar to GRID, TCD-TIMIT consists of recordings in a controlled environment, where the speaker is in front of a plain green wall and their face is evenly illuminated. Compared to GRID, TCD-TIMIT has more speakers, 62 in total (32 males and 30 females, three of which are lipspeakers\footnote{Lipspeakers are professionals trained to make their mouth movements more distinctive. They silently repeat a spoken talk, making lipreading easier for hearing impaired listeners \cite{harte2015tcd}.}), and they pronounce a phonetically balanced group of sentences from the TIMIT corpus.

Other databases have characteristics similar to GRID and TCD-TIMIT (e.g. \cite{patterson2002cuave, sanderson2003noise, abdelaziz2017ntcd, alghamdi2018corpus, anina2015ouluvs2}), but these two are still the most adopted ones, probably for two reasons: the amount of data in them is suitable to train reasonably large deep-learning-based models; their adoption in early AV-SE and AV-SS techniques have made them benchmark datasets for these tasks.

Datasets collected in controlled environments are a good choice for training a prototype designed for a specific purpose or for studying a particular problem. Examples of databases useful in this sense are:  TCD-TIMIT \cite{harte2015tcd} and OuluVS2 \cite{anina2015ouluvs2}, to study the influence of several angles of view;  MODALITY \cite{czyzewski2017audio} and OuluVS2 \cite{anina2015ouluvs2}, to determine the effect of different video frame rates; Lombard GRID \cite{alghamdi2018corpus}, to understand the impact of the Lombard effect, also from several angles of view; RAVDESS \cite{livingstone2018ryerson}, to perform a study of emotions in the context of SE and SS; KinectDigits \cite{schonherr2016environmentally} and MODALITY \cite{czyzewski2017audio}, to determine the importance that supplementary information from the depth modality might have; ASPIRE  \cite{gogate2019cochleanet}, to evaluate the systems in real noisy environments.

\begin{table*}
\caption{{Chronological list of the} main audio-visual speech datasets. The last column indicates the audio-visual speech enhancement and separation articles where the database has been used.}
\centering
\resizebox{0.94\textwidth}{!}{%
\begin{tabular}{l l l l l l l l l}
\toprule
Dataset & Year & \# Speakers & Linguistic Content & Video & Audio & {Additional Characteristics} & AV-SE/SS Papers\\
\midrule
CUAVE \cite{patterson2002cuave}\rule{0pt}{2.5ex}               & 2002 & 36 (19 males) & Connected and isolated & 720$\times$480 & Stereo 44 kHz & Controlled environment & \cite{ephrat2018looking}\\
&&and 20 pairs&digits (7,000 utterances)&29.97 FPS& Mono 16 kHz & Speaker movements\\
&&&&&&Simultaneous speakers\\
%
%
GRID \cite{cooke2006audio}\rule{0pt}{2.5ex}                        & 2006 & 34 (18 males) & Command sentences & 720$\times$576 &50 kHz & Controlled environment & \cite{adeel2019lip, adeel2019contextual, adeel2019novel, arriandiaga2019audio, gabbay2018seeing, gabbay2018visual, gogate2018dnn, gogate2019cochleanet} \\
&&&(1,000 of 3 seconds&25 FPS&& Frontal face & \cite{ideli2019visually, khan2018using, li2020visual, lu2018listen, lu2019audio, 9053033} \\
&&&per speaker)&&&& \cite{sun2020attention, owens2018audio, morrone2019face, pasa2019joined, michelsanti2019training} \\
OuluVS \cite{zhao2009lipreading} \rule{0pt}{2.5ex} & 2009 & 20 (17 males) & 10 everyday greetings& 720$\times$576 & 48 kHz & Controlled environment & $\,$ -- \\
&&&(817 sequences)&25 FPS&&Rotating head movements\\
LDC2009V01 \cite{richie2009audiovisual}\rule{0pt}{2.5ex}   & 2009 & 14 (4 males) & Single words and full & 720$\times$480 & 48 kHz & Controlled environment & \cite{wu2016multi}\\
&&&sentences (7 hours)&29.97 FPS&&Frontal face\\
TCD-TIMIT \cite{harte2015tcd}\rule{0pt}{2.5ex}                    & 2015 & 62 (32 males) & Phonetically rich & 1920$\times$1080 & 48 kHz & Controlled environment & \cite{gabbay2018seeing, gabbay2018visual, ephrat2018looking, gogate2019cochleanet, lu2019audio, morrone2019face}\\
&&3 lipspeakers&sentences (13,826 clips)& 30 FPS&& Straight and 30$^{\circ}$ camera& \cite{pasa2019joined}\\
OuluVS2 \cite{anina2015ouluvs2} \rule{0pt}{2.5ex} & 2015 & 53 (40 males) & Continuous digits & 1920$\times$1080 & 48 kHz & Controlled environment & $\,$ -- \\
&&&and sentences&30 FPS&&Five views\\
&&&&640$\times$480 (front)&&\\
&&&&100 FPS (front)&&\\
KinectDigits \cite{schonherr2016environmentally}\rule{0pt}{2.5ex} & 2016 & 30 (15 males) & English digits 0 -- 9 & 104$\times$80 & Four-channel & Controlled environment & $\,$ -- \\
&&&& 30 FPS&16 kHz  & RGB and depth frames of\\
&&&& & & the mouth region\\
&&&& & & Mic. array recordings\\
LRW \cite{chung2016lip}\rule{0pt}{2.5ex}                             & 2016 & Hundreds & Utterances of 500& 256$\times$256 & 16 kHz & Videos in the wild & \cite{ideli2019visually, ideli2019multi}\\
&&&different words & 25 FPS &  &Recordings from BBC \\
&&&(173 hours)&&&Mostly frontal faces\\
Small Mandarin  Sentences \rule{0pt}{2.5ex}  & 2016 & 1 male & 40 utterances & 320$\times$240 & 48 kHz & Controlled environment & \cite{hou2016audio}\\
Corpus \cite{hou2016audio}&&&(3-4 seconds each)&&& Frontal face &\\
&&&&&& Mandarin &\\
MODALITY \cite{czyzewski2017audio} \rule{0pt}{2.5ex}      & 2017 & 35 (26 males) & Separated commands & 1920$\times$1080 & Eight-channel & Controlled environment & $\,$ -- \\
&&&Continuous sentences& 100 FPS& + phone mic.  & RGB and depth frames \\
&&&(31 hours)& 320$\times$240 (ToF) & 44.1 kHz  & Clean and noisy conditions \\
&&&& 60 FPS (ToF)&  & Mic. array recordings \\
NTCD-TIMIT \cite{abdelaziz2017ntcd}\rule{0pt}{2.5ex}        & 2017 & \multicolumn{5}{l}{Extension of TCD-TIMIT obtained by adding six noise types to the corpus: white, babble, car, living room,} & \cite{sadeghi2019audio, sadeghi2019robust, sadeghi2019mixture, li2020visual}\\
&&street and cafe\\
LRS \cite{chung2016lip2}\rule{0pt}{2.5ex} & 2017 & Several & Continuous sentences & Not specified{$^a$} & Not specified{$^a$} & Videos in the wild & $\,$ -- \\
&&(Not specified{$^a$})&(75.5 hours)&&&Recordings from BBC\\
&&&&&&Mostly frontal faces\\
MV-LRS \cite{chung2017lip}\rule{0pt}{2.5ex}                        & 2017 & Several & Continuous sentences & Not specified{$^a$} & Not specified{$^a$} & Videos in the wild & \cite{afouras2019my}\\
&&(Not specified{$^a$})& (777.2 hours) &&&Recordings from BBC\\
&&&&&&Multiview\\
VoxCeleb \cite{nagrani2017voxceleb}\rule{0pt}{2.5ex}         & 2017  & 1,251 & Continuous sentences & Not specified{$^b$} & Not specified{$^b$} & Videos in the wild from & \cite{owens2018audio}\\
&&(690 males)&(153,516 utterances,&&&Youtube\\
&&&352 hours)&&&Challenging multi-speaker\\
&&&&&&acoustic environments\\
Mandarin Sentences \rule{0pt}{2.5ex}  & 2018 & 1 male & 320 utterances & 1920$\times$1080 & 48 kHz & Controlled environment & \cite{hou2017audio, gabbay2018visual, ephrat2018looking, gogate2019cochleanet}\\
Corpus \cite{hou2017audio}&&&(3-4 seconds each)&30 FPS&& Frontal face &\\
&&&&&& Mandarin &\\
Obama Weekly Addresses \cite{gabbay2018visual}\rule{0pt}{2.5ex}  & 2018 & 1 male & Continuous sentences& Not specified{$^c$} & Not specified{$^c$} & Wide variety of lighting, & \cite{gabbay2018visual}\\
&&&(300 videos of 2-3&&&face pose, background,\\
&&&minutes long)&&&scaling and audio\\
&&&&&&recording conditions\\
Lombard GRID \cite{alghamdi2018corpus}\rule{0pt}{2.5ex} & 2018 & 54 (24 males) & Command sentences & 720$\times$480 (front) & 48 kHz & Controlled environment & \cite{michelsanti2019effects, michelsanti2019deep}\\
&&&(50 Lombard and 50&24 FPS (front)&&Frontal face\\
&&& plain per speaker)&864$\times$480 (side)&&Lombard effect recordings\\
&&&&30 FPS (side)&&Straight and side camera\\
RAVDESS \cite{livingstone2018ryerson}\rule{0pt}{2.5ex} & 2018 & 24 actors & Continuous sentences & 1920$\times$1080 & 48 kHz & Controlled environment & $\,$ -- \\
&& (12 males) & and songs (2452  & 30 FPS && Emotional speech and  \\
&&&audio-visual clips)&&& singing at two levels\\
&&&&&& of intensity\\
VoxCeleb2 \cite{chung2018voxceleb2}\rule{0pt}{2.5ex}       & 2018 & 6,112 & Continuous sentences & Not specified{$^b$} & Not specified{$^b$} & Videos in the wild from & \cite{afouras2018conversation, joze2019mmtm, luo2019audio, 9054180, iuzzolino2020av, chung2020facefilter}\\
&&(3,761 males)&(1,128,246 utterances,&&&Youtube \\
&&&2,442 hours)&&&Challenging visual and\\
&&&&&&auditory environments\\
LRS2 \cite{afouras2018deep}\rule{0pt}{2.5ex}                     & 2018  & Hundreds & Continuous sentences & Not specified{$^b$} & Not specified{$^b$} & Videos in the wild & \cite{afouras2018conversation, afouras2019my, wu2019time, ideli2019multi, 9054180}\\
&&&(up to 100 characters&&&Recordings from BBC \\
&&&each -  224.5 hours)\\
LRS3 \cite{afouras2018lrs3}\rule{0pt}{2.5ex}                       & 2018  & Around 5,000 & Continuous sentences & 224$\times$224 & 16 kHz & Videos from TED and & \cite{afouras2019my, ochiai2019multimodal, qu2020multimodal}\\
&&& (438 hours) & 25 FPS &&TEDx YouTube channels\\
AVSpeech \cite{ephrat2018looking}\rule{0pt}{2.5ex}           & 2018 & 150,000 & Continuous sentences  & Not specified{$^b$} & Not specified{$^b$} & Videos in the wild & \cite{ephrat2018looking, inan2019evaluating}\\
&&&(4,700 hours)&&&Wide variety of people,\\
&&&& & & languages and face poses\\
AV Chinese Mandarin \cite{tan2019audio}\rule{0pt}{2.5ex} & 2019 & Several & Continuous sentences & Not specified{$^b$} & Not specified{$^b$} & Mandarin lectures from  & \cite{gu2020multi, tan2019audio, xu2020neural}\\
&& (Not specified)&(155 hours)&& &YouTube\\
&&&&& &Grayscale frames of lips\\
AVA-ActiveSpeaker \cite{gu2018ava, roth2019ava}\rule{0pt}{2.5ex} & 2019 & Several & Continuous sentences & Not specified{$^b$} & Not specified{$^b$} & Movie and TV videos from&  $\,$ -- \\
&& (Not specified)&(38.5 hours)&& &YouTube\\
&&&&&& Human-labelled frames\\
ASPIRE \cite{gogate2019cochleanet}\rule{0pt}{2.5ex} & 2019 & 3 (1 male) & Command sentences & 1920$\times$1080 & 44.1 kHz & Recordings in real noisy &  \cite{gogate2019cochleanet, gogate2019av} \\
&&&(6,000 utterances)&30 FPS& Binaural & places and isolated booth\\
\bottomrule
 \multicolumn{7}{l}{$^a$We could not get this information because the database is not available to the public due to license restrictions. \rule{0pt}{2.5ex} }\\
 \multicolumn{7}{l}{$^b$Since the material is from YouTube, we can expect variable video resolution and audio sample rate.}\\
 \multicolumn{7}{l}{$^c$Weekly addresses from The Obama White House YouTube channel. Original video resolution of 1920$\times$1080 at 30 FPS and audio sample rate of 44.1 kHz.}
\end{tabular}}
\label{tab:databases}
\end{table*}

\subsection{{Data in the Wild}}

More recently, an effort of the research community has been put to gather \emph{data in the wild}, in other words recordings from different sources without the careful planning and setting of the controlled environment used in conventional datasets, like the already mentioned GRID and TCD-TIMIT. The goal of collecting such large-scale datasets, characterised by a vast variety of speakers, sentences, languages and visual/auditory environments, not necessarily in a controlled lab setup, is to have data that resemble real-world recordings. One of the first AV speech in-the-wild databases is LRW \cite{chung2016lip}. LRW was specifically collected for VSR and consists of around 170 hours of AV material from British television programs. The utterances in the dataset are spoken by hundreds of speakers and are divided into 500 classes. Each sentence of a class contains a non-isolated keyword between 5 and 10 characters. The trend of collecting larger datasets has continued in subsequent collections which consist of materials from British television programs \cite{chung2016lip2, chung2017lip, afouras2018deep}, generic YouTube videos \cite{nagrani2017voxceleb, chung2018voxceleb2}, TED talks \cite{afouras2018lrs3, ephrat2018looking}, movies \cite{roth2019ava} or lectures \cite{tan2019audio, ephrat2018looking}. Among them, AVSpeech is the largest dataset used for AV-SE and AV-SS, with its 4,700 hours of AV material. It consists of a wide range of speakers (150,000 in total), languages (mostly English, but also Portuguese, Russian, Spanish, German and others) and head poses (with different pan and tilt angles). Each video clip contains only one talking person and does not have acoustic background interferences. 

The large-scale in-the-wild databases, as opposed to the ones containing recordings in controlled environments, are particularly suitable for training deep models that must perform robustly in real-world situations. {However, although some in-the-wild datasets are more used than others (as can be seen in the last column of Table} \ref{tab:databases}), {there is not a standard benchmark dataset used to perform the experiments in unconstrained conditions.}


\section{ {Performance Assessment}} \label{sec:evaluation}

The main aspects generally of interest for SE and SS are \emph{quality} and \emph{intelligibility}. Speech quality is largely subjective \cite{loizou2013speech, deller2000discrete} and can be defined as the result of the judgement based on the characteristics that allow to perceive speech according to the expectations of a listener \cite{jekosch2006voice}. Given the high number of dimensions that the quality attribute possesses and the different subjective concept of what is high and low quality for every person, a large variability is usually observed in the results of speech quality assessments \cite{loizou2013speech}. 
On the other hand, intelligibility can be considered a more objective attribute, because it refers to the speech content \cite{loizou2013speech}. Still, a variability in the results of intelligibility assessments can be observed due to the individual speech perception, which has an impact on the ability of recognising words and/or phonemes in different situations.

In the rest of this Section, we review how AV-SE and AV-SS systems are evaluated, with a particular focus on speech quality and speech intelligibility. A summary of the different methods and measures used in the literature are shown in Table~\ref{tab:metrics}.

\subsection{Listening Tests}

A proper assessment of SE and SS systems should be conducted on the actual receiver of the signals. In many scenarios (e.g. hearing assistive devices, teleconferences etc.), the receiver is a human user and \emph{listening tests}, performed with a population of the expected end users, are the most reliable way for the evaluation.

The tests that are currently employed for the assessment of AV-SE and AV-SS systems are typically adopted from the AO domain, i.e. they follow procedures validated for AO-SE and AO-SS techniques. Although different kinds of listening tests exist, some general recommendations include:
\begin{itemize}
	\item Several subjects are required to be part of the assessment. The number depends on the task, the listeners' experience and the magnitude of the performance differences (e.g. between a system under development and its predecessor) that one wishes to detect. Generally, fewer subjects are required, if they are expert listeners.
	\item Before the actual test, a training phase allows the subjects to familiarise themselves with the material and the task.
	\item The speech signals are presented to the listeners in a random order.
	\item To reduce the impact of listening fatigue, long test sessions are avoided.
\end{itemize}

The most common method used in SE \cite{loizou2013speech} to assess speech quality is the \emph{mean opinion score} (MOS) test \cite{rec1990bs562, rec2019bs1284_2, rec1996p830}. This test is characterised by a five-point rating scale (cf. the `OVRL' column of Table \ref{tab:mos}) and was adopted in three AV works \cite{adeel2019lip, adeel2019contextual, adeel2019novel}. However, the MOS scale was originally designed for speech coders, which introduce different distortions than the ones found in SE \cite{loizou2013speech}. Therefore, an extended standard \cite{rec2003p835} was proposed and five-point discrete scales were used to rate not only the overall (OVRL) quality (like in the MOS test), but also the signal (SIG) distortion and the background (BAK) noise intrusiveness (cf. Table~\ref{tab:mos}). This kind of assessment was adopted to evaluate the AV system in~\cite{hou2017audio}.

\begin{table}
\caption{Signal (SIG), background (BAK) and overall (OVRL) quality rating scales according to \cite{rec2003p835}. The overall quality scale is the same as the mean opinion score scale.}
\centering
\resizebox{0.48\textwidth}{!}{%
\begin{tabular}{c l l l}
\toprule
Rating & SIG & BAK & OVRL\\
\midrule
5 & Not distorted & Not noticeable & Excellent  \\
4 & Slightly distorted & Slightly noticeable & Good \\
3 & Somewhat distorted & Noticeable but not intrusive & Fair \\
2 & Fairly distorted & Somewhat intrusive & Poor \\
1 & Very distorted & Very intrusive & Bad \\
\bottomrule
\end{tabular}}
\label{tab:mos}
\end{table}

\begin{table*}
\caption{{List of the} main performance assessment methods for audio-visual speech enhancement and separation. The last column indicates the audio-visual speech enhancement and separation articles where the evaluation method has been used. 
}
\centering
\resizebox{0.94\textwidth}{!}{%
\begin{tabular}{l l l l l}
\toprule
Type & Evaluation Method & Year & {Main Characteristics} & AV-SE/SS papers\\
\midrule
Listening tests for speech & MOS \cite{rec1990bs562, rec2019bs1284_2, rec1996p830} & $\,$ --  & Audio-only listening test with 5-point& \cite{adeel2019lip, adeel2019contextual, adeel2019novel}\\
 quality assessment&&& rating scale\\
& SIG / BAK / OVRL \cite{rec2003p835} \rule{0pt}{2.5ex} & 2003 & Extension of MOS considering signal & \cite{hou2017audio}\\
&&& distorsion and noise intrusiveness\\
 & MUSHRA \rule{0pt}{2.5ex} \cite{itu2003bs1534_1} & 2003  & Audio-only listening test with & \cite{gogate2019cochleanet, gogate2019av}\\
 &&& continuous rating scale\\
 & MUSHRA-like audio-visual  \rule{0pt}{2.5ex} & 2019 & MUSHRA test using audio-visual & \cite{michelsanti2019deep}\\
 & test  \cite{itu2003bs1534_1, michelsanti2019deep} && stimuli\\
 \hdashline
Listening tests for speech \rule{0pt}{2.5ex}&DRT \cite{voiers1983evaluating} & 1983 & Audio-only listening test using & $\,$ -- \\
intelligibility assessment &&& rhyming words\\
 &HINT \cite{nilsson1994development}  \rule{0pt}{2.5ex}& 1994 & Audio-only listening test using & $\,$ --\\
 &&& everyday sentences\\
 & Matrix-like audio-visual  \rule{0pt}{2.5ex} & 2019 & Matrix test using audio-visual& \cite{michelsanti2019deep}\\
&  test \cite{michelsanti2019deep}&&  stimuli \cite{alghamdi2018corpus} \\
\midrule
Estimators of speech quality &PESQ \cite{rix2001perceptual, itu2001862, rec2003p, rec2005p}  \rule{0pt}{2.5ex} & 2001 & Designed to assess quality across a & \cite{adeel2019lip, adeel2019contextual, adeel2019novel, afouras2018conversation, aldeneh2020self, arriandiaga2019audio, chuang2020lite, ephrat2018looking, gabbay2018seeing} \\ 
based on perceptual models&&& wide range of codecs and network & \cite{gabbay2018visual, gogate2019cochleanet, gogate2018dnn,  gu2020multi, hou2017audio, ideli2019multi, ideli2019visually} \\
&&&conditions mostly for telephony & \cite{inan2019evaluating, iuzzolino2020av, joze2019mmtm, khan2018using, li2020visual, 9054180 } \\
&&&&\cite{michelsanti2019training, michelsanti2019effects,  michelsanti2019deep, morrone2019face, sadeghi2019audio, sadeghi2019robust, sadeghi2019mixture}\\
&&&&\cite{ sun2020attention, tan2019audio, wu2016multi, 9053033, xu2020neural}\\
&CSIG / CBAK / COVRL  \cite{hu2007evaluation}   \rule{0pt}{2.5ex}& 2007 & Composite measures which combine& \cite{ideli2019visually}\\
&&& basic objective measures \\
&HASQI \cite{kates2010hearing, kates2014hearing2}  \rule{0pt}{2.5ex}& 2010 & Specifically designed for hearing- & \cite{hou2016audio, hou2017audio}\\
&&&impaired listeners\\
&POLQA \cite{rec2011p}  \rule{0pt}{2.5ex}& 2011 & PESQ successor & $\,$ -- \\
&ViSQOL  \cite{hines2012visqol, hines2015visqol}   \rule{0pt}{2.5ex}& 2012 & Specifically designed for voice over& \cite{ephrat2018looking, morrone2019face}\\
&& &IP transmission\\
\hdashline
Estimators of speech quality &SNR   \rule{0pt}{2.5ex} & $\,$ -- & It does not provide a proper & \cite{gabbay2018seeing, gabbay2018visual, inan2019evaluating, aldeneh2020self}\\
based on energy ratios &(Signal-to-Noise Ratio)&&estimation of speech distortion\\
&SSNR / SSNRI  \rule{0pt}{2.5ex}& $\,$ -- & Assessment of short-time & \cite{ideli2019visually, sun2020attention, hou2016audio}\\
&(Segmental SNR) &&behaviour&\\
&(SSNR Improvement)\\
 & SDI \cite{chen2006new} \rule{0pt}{2.5ex} & 2006 & It provides a rough distortion & \cite{hou2016audio, hou2017audio}\\
&&&measure\\
&SDR \cite{vincent2006performance}  \rule{0pt}{2.5ex}& 2006 & Specifically designed for blind audio & \cite{afouras2019my, afouras2018conversation, arriandiaga2019audio, chung2020facefilter, ephrat2018looking, gabbay2018seeing, gu2020multi}\\
&&&source separation&\cite{ideli2019visually, ideli2019multi, inan2019evaluating, khan2018using, 9054180, li2020visual, luo2019audio}\\
&&&& \cite{lu2018listen, lu2019audio, ochiai2019multimodal, owens2018audio, morrone2019face, pasa2019joined}\\
&&&& \cite{qu2020multimodal, sadeghi2019audio, sadeghi2019robust, sadeghi2019mixture}\\
&SIR \cite{vincent2006performance}  \rule{0pt}{2.5ex}& 2006 & Specifically designed for blind audio & \cite{gabbay2018seeing, afouras2018conversation, lu2018listen, lu2019audio, ideli2019multi, khan2018using}\\
&&&source separation&\cite{owens2018audio}\\
&SAR \cite{vincent2006performance} \rule{0pt}{2.5ex}& 2006 & Specifically designed for blind audio & \cite{gabbay2018seeing, owens2018audio, lu2018listen, lu2019audio, ideli2019multi, khan2018using}\\
&&&source separation\\
&SI-SDR \cite{le2019sdr}  \rule{0pt}{2.5ex}& 2019 & Extension of SDR to make it scale- & \cite{gogate2019cochleanet, ideli2019visually, gu2020multi, wu2019time, tan2019audio}\\
&&&invariant\\
\hdashline
Estimators of speech &SII \cite{american1997american} \rule{0pt}{2.5ex} & 1997 & Used for additive stationary noise or & \cite{ideli2019visually}\\
intelligibility&& & bandwidth reduction\\
&CSII \cite{kates2004coherence}  \rule{0pt}{2.5ex}& 2004 & Extension of SII for broadband peak- & \cite{ideli2019visually}\\
&& & clipping and center-clipping distortion\\
&ESII \cite{rhebergen2005speech}  \rule{0pt}{2.5ex}& 2005 & Extension of SII for fluctuating noise & \cite{ideli2019visually}\\
&STOI \cite{taal2011algorithm}  \rule{0pt}{2.5ex}& 2011 & Able to predict quite accurately speech  & \cite{ephrat2018looking, gogate2019cochleanet, ideli2019visually, inan2019evaluating, afouras2018conversation, chuang2020lite, gu2020multi}\\
&&&intelligibility in several situations& \cite{iuzzolino2020av, joze2019mmtm, khan2018using, hou2017audio}\\
&HASPI \cite{kates2014hearing} \rule{0pt}{2.5ex}& 2014 & Specifically designed for hearing- & \cite{hou2016audio, hou2017audio}\\
&&&impaired listeners \\
&ESTOI \cite{jensen2016algorithm} \rule{0pt}{2.5ex}& 2016 & Extension of STOI for highly & \cite{tan2019audio, michelsanti2019training, michelsanti2019effects, michelsanti2019deep, ideli2019visually, ideli2019multi}\\
&&& modulated noise sources \\
\hdashline
Automatic speech recognition &WER \rule{0pt}{2.5ex} & $\,$ --  & Word-level comparison & \cite{afouras2018conversation, afouras2019my, li2020visual, gu2020multi, qu2020multimodal, xu2020neural} \\ 
performance&(Word Error Rate)&\\
&PER  \rule{0pt}{2.5ex}& $\,$ -- & Phone-level comparison & \cite{pasa2019joined}\\
&(Phone Error Rate)&\\
\midrule
Computational efficiency &RTF & $\,$ -- & Ratio between GPU processing & \cite{gu2020multi}\\
&(Real-Time Factor)&& time and audio time. \\
\bottomrule
\end{tabular}}
\label{tab:metrics}
\end{table*}

A distinct quality assessment procedure, the \emph{multi stimulus test with hidden reference and anchor} (MUSHRA) \cite{itu2003bs1534_1}, was used in \cite{gogate2019cochleanet, gogate2019av, michelsanti2019deep}. In this case, the listeners are presented with speech signals to be rated using a continuous scale from 0 to 100, consisting of 5 equal intervals labelled as `bad', `poor', `fair', `good', and `excellent'. The test is divided into several sessions. In each session, the subjects are asked to rate a fixed number of signals under test (processed and/or noisy speech signals), one hidden reference (the underlying clean speech signal) and at least one hidden anchor (a low-quality version of the reference). In addition, the clean speech signal (i.e. the unhidden reference) is provided. The hidden reference allows to understand whether the subject is able to detect the artefacts of the processed signals, while the hidden anchor provides a lowest-quality fixed-point in the MUSHRA scale, determining the dynamic range of the test. Having the possibility to switch among the signals at will, the listeners can make comparisons with a high degree of resolution.

Together with speech quality, also intelligibility should be assessed with appropriate listening tests. It is possible to group the intelligibility tests into three classes, based on the speech material adopted \cite{loizou2013speech}:
\begin{itemize}
	\item \emph{Nonsense syllable tests}  - Listeners need to recognise nonsense syllables drawn from a list \cite{fletcher1929articulation, miller1955analysis}. Usually, it is hard to build such a list of syllables where each item is equally difficult to be identified by the subjects, hence these tests are not very common.
	\item \emph{Word tests} - Listeners are asked to identify words drawn from a phonetically balanced list \cite{egan1948articulation} or rhyming words \cite{fairbanks1958test, house1965articulation, voiers1983evaluating}. Among these tests, the \emph{diagnostic rhyme test} (DRT) \cite{voiers1983evaluating} is extensively adopted to evaluate speech coders \cite{loizou2013speech}. The main criticism about word tests is that they may be unable to predict the intelligibility in real-world scenarios, where a listener is usually exposed to sentences, not single words.
	\item \emph{Sentence tests} - Listeners are presented with sentences and are asked to identify keywords or recognise the whole utterances. It is possible to distinguish these tests between the ones that use everyday sentences \cite{kalikow1977development, nilsson1994development} and the ones that use sentences with a fixed syntactical structure, known as \emph{matrix tests} \cite{hagerman1982sentences, Wagener1999a, Wagener1999b, Wagener1999c, ozimek2010polish, hochmuth2012spanish}. One of the most commonly used sentence tests is the \emph{hearing in noise test} (HINT) \cite{nilsson1994development}, also adapted for different languages, including Canadian-French \cite{vaillancourt2005adaptation}, Cantonese \cite{wong2005development}, Danish \cite{nielsen2009development, nielsen2011danish} and Swedish \cite{hallgren2006swedish}. 
\end{itemize}
A simple way to quantify the intelligibility for the previously mentioned tests is by calculating the so-called \emph{percentage intelligibility} \cite{loizou2013speech}. This measure indicates the percentage of correctly identified syllables, words or sentences at a fixed SNR. The main drawback is that it might be hard to find the SNR at which the test can be optimally performed, because floor or ceiling effects might occur if the listeners' task is too hard or too easy. This issue can be mitigated by testing the system at several SNR within a pre-determined range, at the expense of the time needed to conduct the listening experiments. As an alternative, speech intelligibility can be measured in terms of the so-called \emph{speech reception threshold} (SRT), which is the SNR at which listeners correctly identify the material they are exposed to with a 50\% accuracy\footnote{Variants exist where a different percentage is used.} \cite{loizou2013speech}. The SRT is determined with an adaptive procedure, where the SNR of the presented stimuli increases or decreases by a fixed amount at every trial based on the subject's previous response. In this case, the main drawback is that the test is not informative for SNRs that substantially differ from the determined SRT.

Speech intelligibility tests are yet to be adopted by the AV-SE and AV-SS community. In fact, an intelligibility evaluation involving human subjects for AV-SE can only be found in \cite{michelsanti2019deep}. There, listeners were exposed to speech signals from the Lombard GRID corpus \cite{alghamdi2018corpus} processed with several systems and were asked to determine three keywords in each sentence. The results were reported in terms of percentage intelligibility for four different SNRs distributed  in uniform steps between $-$20~dB and $-$5~dB.

An important element to consider is the modality in which the stimuli are presented. The listening tests conducted in AV-SE works, like \cite{hou2017audio, gogate2019cochleanet, gogate2019av, adeel2019lip, adeel2019contextual}, generally used AO signals. Although simpler to conduct, this kind of setting has the disadvantage of completely ignoring the visual information, which has an impact on speech perception \cite{sumby1954visual, mcgurk1976hearing}. Moreover, it is important to perform tests under the same conditions in which the systems are used in practice. In the situation of human-receiver devices, the user is often exposed to both auditory as well as visual stimuli. Consequently, such systems ought to be tested in natural conditions with human subjects receiving both auditory and corresponding visual stimuli. This is the reason why the tests in \cite{michelsanti2019deep} were performed with AV signals. However, an AV setup entails a challenging interpretation of the results due to several factors, as highlighted in \cite{hussain2017towards, michelsanti2019deep}:
\begin{itemize}
\item There is a big difference among individuals in lip-reading abilities. This difference is not reflected in the variation in auditory perception skills \cite{summerfield1992lipreading}.
\item The per-subject fusion response to discrepancies between the auditory and the visual syllables is large and unpredictable \cite{mallick2015variability}.
\item The availability of visual information makes ceiling effects more probable to occur.
\end{itemize}
These considerations suggest a strong need for exploration and development of ecologically valid paradigms for AV listening tests \cite{hussain2017towards}, which should reduce the variability of the results and provide a robust and reliable estimation of the performance in real-world scenarios. A first step towards achieving this goal is to perform tests in which the subjects are carefully selected within a homogeneous group and exposed to AV speech signals that resemble actual conversational settings from a visual and an acoustic perspective.

\subsection{Objective Measures}

Listening tests are ideal in the assessment performance of SE and SS systems. However, conducting such tests can be time consuming and costly \cite{loizou2013speech}, in addition to requiring access to a representative group of end users. Therefore, researches developed algorithmic methods for repeatable and fast evaluation, able to estimate the results of listening tests without listening fatigue effects. Such methods are often called \emph{objective measures} and most of them exploit the knowledge from low-level (e.g. psychoacoustics) and high-level (e.g. linguistics) human processing of speech \cite{loizou2013speech} (cf. Table~\ref{tab:metrics}).

The most widely used objective measure to assess speech quality for AV-SE and AV-SS is the \emph{perceptual evaluation of speech quality} (PESQ) measure \cite{rix2001perceptual, itu2001862, rec2003p, rec2005p}. PESQ was originally designed for telephone networks and codecs. It is a fairly complex algorithm consisting of several components, including level equalisation, pre-processing filtering, time alignment, perceptual filtering, disturbance processing and time averaging. All these steps are used to take into account relevant psychoacoustic principles:
\begin{itemize}
\item The frequency resolution of the human auditory system is not uniform, showing a higher discrimination for low frequencies \cite{stevens1937scale}.
\item Human loudness perception is not linear, meaning that the ability to perceive changes in sound level varies with frequency \cite{zwicker2013psychoacoustics}.
\item Masking effects might hinder the perception of weak sounds \cite{gelfand2016hearing}.
\end{itemize}
The output of PESQ is supposed to approximate the MOS score and it is a value generally ranging between 1 and 4.5, although a lower score can be observed for extremely distorted speech signals. Rix et al. \cite{rix2001perceptual} reported a high correlation with listening tests in several conditions, i.e. mobile, fixed, voice over IP (VoIP) and multiple type networks. A later study \cite{hu2007evaluation} showed that PESQ correlates well also with the overall quality of signals processed with common SE algorithms.

As new network and headset technologies were introduced, PESQ was not able to accurately predict speech quality. Therefore, a new measure, the perceptual objective listening quality assessment (POLQA) \cite{rec2011p}, was introduced. POLQA is considered the successor of PESQ and it is particularly recommended in scenarios where its predecessor performs poorly or cannot be used, e.g. for high background noise, super-wideband speech, variable delay and time scaling. Although POLQA correlates well with listening test results, outperforming PESQ \cite{rec2011p, beerends2013perceptual}, it has not been used to evaluate AV-SE and AV-SS systems yet.

For SS techniques, assessing the overall quality of the processed signals might not be sufficient, because it is desirable to have measures that characterise different speech quality degradation factors. For this reason, the majority of AV-SS systems are evaluated using a set of measures contained in the \emph{blind source separation} (BSS) \emph{Eval} toolkit \cite{vincent2006performance}. The computation of these measures consists of two steps. First, each of the processed signals is decomposed into four terms, representing the components perceived as coming from: the desired speaker, other target speakers (generating cross-talk artefacts), noise sources and other causes (e.g. processing artefacts). The second step provides performance criteria from the computation of energy ratios related to the previous four terms: \emph{source to distortion ratio} (SDR), \emph{source to interferences ratio} (SIR), \emph{sources to noise ratio} and \emph{sources to artefacts ratio} (SAR). Although a reasonable correlation was found between SIR and human ratings of interference \cite{ward2018bss}, other experiments \cite{cano2016evaluation, ward2018bss} showed that energy-based measures are not ideal for determining perceptual sound quality for SS algorithms.

Besides speech quality estimators, objective intelligibility measures have also been developed. Among them, the \emph{short-time objective intelligibility} (STOI) measure \cite{taal2011algorithm} is the most commonly used for AV-SE and AV-SS. STOI is based on the computation of a correlation coefficient between the short-time overlapping temporal envelope segments of the clean and the degraded/processed speech signals. It has been shown that STOI correlates well with the results of intelligibility listening experiments \cite{taal2011algorithm, falk2015objective, xia2012evaluation}. An extension of STOI, ESTOI, was later proposed \cite{jensen2016algorithm} to provide a more accurate prediction of speech intelligibility in presence of highly modulated noise sources.

Table \ref{tab:metrics} indicates also other measures that we have not presented above, because they are less adopted in AV-SE and AV-SS works. However, it is worth mentioning some of them, since they can be used by researchers to evaluate the systems for specific purposes. For example, the \emph{hearing-aid speech quality index} (HASQI) \cite{kates2010hearing, kates2014hearing2} and the \emph{hearing-aid speech perception index} (HASPI) \cite{kates2014hearing} are two measures that have been specifically designed to evaluate speech quality and and intelligibility as perceived by hearing-impaired listeners. Sometimes, the evaluation of a system is expressed in terms of word error rate (WER) as measured by an ASR system (cf. Table \ref{tab:metrics}). This measure assumes that the receiver of the signals is a machine, not a human, and it provides additional performance information for specific applications, e.g. video captioning for teleconferences or augmented reality.

Most of the objective measures used to evaluate AV-SE and AV-SS systems have two main limitations to be desirably addressed in future works. First, they require the target speech signal(s) in order to produce a quality or an intelligibility estimate of the degraded/processed signal(s). These measures are known as \emph{intrusive estimators}. For algorithm development, where clean speech reference is readily available, this assumption is reasonable. However, for in-the-wild tests, it is not possible to collect reference signals and intrusive estimators cannot be adopted.

The other limitation is the use of AO signals in all the objective measures. As already pointed out for listening tests, ignoring the visual component of speech may cause an erroneous estimation of the system performance in many real-world situations, where the listener is able to look at the speaker. In order to develop new predictors of quality and intelligibility in an AV context, a substantial amount of data from AV listening tests is required. When such AV data is available, it would be possible to understand the factors influencing human AV perception of processed speech and properly design and validate new objective measures.

\subsection{Beyond Speech Quality and Intelligibility}

When considering SE and SS systems, aspects other than speech quality and intelligibility might be of interest to assess. Some systems, like hearing assistive devices and teleconference systems, have a low-latency requirement, because they need to deliver processed signals to allow real-time conversations. In this case, it might be relevant to report a measure of the \emph{computational efficiency} of the approach under analysis. An example is the so-called real-time factor (RTF), used in \cite{gu2020multi} and defined as the ratio between the processing time and the duration of the signal.

Sometimes, a given processed speech signal could be fully intelligible, but the effort that the listener must put into the listening task could be substantial in order to be able to understand the speech content. Therefore, it might be important to measure the energy that a subject needs to invest in a listening task, i.e. the \emph{listening effort}. As for speech quality and intelligibility, the listening effort may be measured with listening tests \cite{feuerstein1992monaural, zekveld2010pupil}.

Moreover, speech carries a lot of additional information, e.g. about the speaker, including \emph{gender}, \emph{age}, \emph{emotional state}, \emph{mood}, their \emph{location} in the scene, etc. These aspects might be important and SE or  SS systems should ideally preserve them even after the processing of a heavily corrupted speech signal (cf. \cite{han2020real}, in which the proposed system is specifically tested for its ability to preserve spatial cues). Standardised methods for the assessment of these aspects of AV speech are currently lacking, but they would be important to develop in order to guarantee high performance to the end users.




\section{Conclusion} \label{sec:conclusion}

In this paper, we presented an overview of deep-learning-based approaches for audio-visual speech enhancement (AV-SE) and audio-visual speech separation (AV-SS). {As expected, visual information provides a benefit for both speech enhancement and separation. In particular, AV-SE systems either outperform their audio-only counterpart for very low signal-to-noise ratios (SNRs) or show similar performance at high SNRs. Performance improvements can be seen across all visemes, with better results for sounds easier to be distinguished visually} \cite{aldeneh2020self}. {Regarding speech separation, audio-visual systems not only outperform their audio-only counterpart, but, since vision is a strong guidance, they are also unaffected by the source permutation problem, occurring when the separated speech signals are assigned inconsistently to the sources.}

{Throughout the paper,} {we surveyed a large number of approaches, deliberately avoiding to advocate a method over another based on their performance. This choice was motivated by the fact that a fair comparison of AV-SE and AV-SS approaches is hard, given the wide range of possible applications, each of them having different requirements (e.g. regarding latency, computational complexity and real-time processing). In addition, the lack of standardised audio-visual evaluation procedures, either in the form of listening tests or objective measures, makes the results obtained from such a comparison hardly interpretable and not representative of actual real-world conditions. The design of an audio-visual evaluation framework (cf. the proposal in} \cite{hussain2017towards}) would be extremely valuable for the community, but it is clearly outside of the scope of this overview. Instead, we leave the reader to decide which methods to use and further investigate, based on the provided description and discussion of the main elements characterising state-of-the-art systems, namely: acoustic features; visual features; deep learning methods; fusion techniques; training targets and objective functions.

We saw that AV-SE and AV-SS systems generally use the short-time magnitude spectrogram as acoustic input. Since the short-time magnitude spectrogram is not a complete representation of the acoustic signal, some methods exploit the phase information, the complex spectrogram or directly the time-domain signal. Regarding the visual input, although raw data is often used, low-dimensional features are preferred in several works. This choice serves two purposes: first, it allows to reduce the complexity of AV-SE and AV-SS algorithms, since the dimensionality of the data to process is lower; secondly, it makes it possible to train deep learning models for AV-SE and AV-SS with less data, because the low-dimensional features usually adopted are already somewhat robust to several factors, such as illumination conditions, face poses, etc. Future systems, where development data and computational resources may be abundant, might aim for end-to-end training using directly raw visual and acoustic signals as input.


In state-of-the-art AV-SE and AV-SS systems, the actual data processing is obtained with deep-learning-based techniques. Generally, acoustic and visual features are processed separately using two neural network models. Then, the output vectors of these models are fused, often by concatenation, and, afterwards, used as input to another deep learning model. This strategy is convenient, because it is very easy to implement. However, it comes with a major drawback: a simple concatenation does not allow to control how the information from the acoustic and the visual modalities is treated. As a consequence, one of the two modalities may dominate over the other, determining a decrease in the system's performance. Among the strategies adopted to tackle this problem, attention-based mechanisms, which allow the systems to attend to relevant parts of the input, mitigate the potential unbalance caused by concatenation-based fusion.

The last two elements of AV-SE and AV-SS systems are training targets, i.e. the desired output of a deep learning model, and objective functions, i.e. functions that measure the distance between the desired output of a model and its actual output. Although a few approaches tried to directly approximate the target speech signal(s) in time domain, more often a time-frequency (TF) representation of the signals is used. In particular, the deep-learning-based systems are generally trained to minimise the mean squared error (MSE) between the network output and the (potentially transformed) TF coefficients of the training target, which can be either the clean magnitude spectrogram or a mask that is applied to the noisy spectrogram to obtain an enhanced speech signal. Among the two training targets, the latter is usually preferred, because a mask has been empirically found to be easier to estimate with deep learning if compared to the clean magnitude spectrogram.

We also presented three other aspects related to AV-SE and AV-SS, since they can provide additional insights. First, we described deep-learning-based methods used to solve two related tasks: speech reconstruction from silent videos and audio-visual sound source separation for non-speech signals. In particular, we reported a chronological evolution of these fields, because they influenced the first AV-SE and AV-SS approaches and they may still provide a source of inspiration for AV-SE and AV-SS research and vice versa.

Second, we surveyed audio-visual speech datasets, since data-driven methods, like the ones based on deep learning, heavily rely on them. We saw that AV-SE and AV-SS research can still benefit from data collected in a controlled environment to study specific phenomena, like Lombard effect. The general tendency, however, is to use large-scale in-the-wild datasets to make the deep-learning-based systems robust to the variety of conditions that may be present in real-world applications. 

Third, we reviewed the principal methodologies used to assess the performance of AV-SE and AV-SS. Specifically, we considered listening tests and objective measures. The former represent the ideal way to assess processed speech signals and must be employed eventually for a realistic system evaluation. However, they are generally time-consuming and costly to conduct. The latter allow to estimate some speech aspects, like quality and intelligibility, in a quick and easily repeatable way, which is highly desirable in the development phase of audio-visual systems. Although many objective measures exist, it might be reasonable to choose the ones that are widely adopted, to make comparisons with previous approaches, and that are reported to correlate well with listening test results. Examples include PESQ, STOI (or its extended version, ESTOI), and SDR (or its scale-invariant definition, SI-SDR). Currently, such objective measures are audio-only. This is in contrast to human communication, which is generally audio-visual.

Finally, we identified several future research directions. Some of them address aspects, such as robustness to a variety of acoustic and visual conditions, to be applied e.g. in teleconferences, and reduction of the computational complexity of deep learning algorithms, especially relevant for low-resource devices like hearing aids. Others, like the investigation of new paradigms for audio-visual fusion, are more focused on a better exploitation of properties and constraints in multimodal systems, and they could, for example, further affect audio-visual speech recognition, audio-visual emotion recognition and audio-visual temporal synchronisation.

\section*{Acknowledgment}

{The authors would like to thank Laurent Girin and the anonymous reviewers for their comments, which helped in improving the overall quality of the article.}

This research is partially funded by the William Demant Foundation.

\ifCLASSOPTIONcaptionsoff
  \newpage
\fi



%

\bibliographystyle{ieeetranS}
\bibliography{av}

%

\vfill
\begin{IEEEbiography}[{\includegraphics[width=1in,height=1.25in,clip,keepaspectratio]{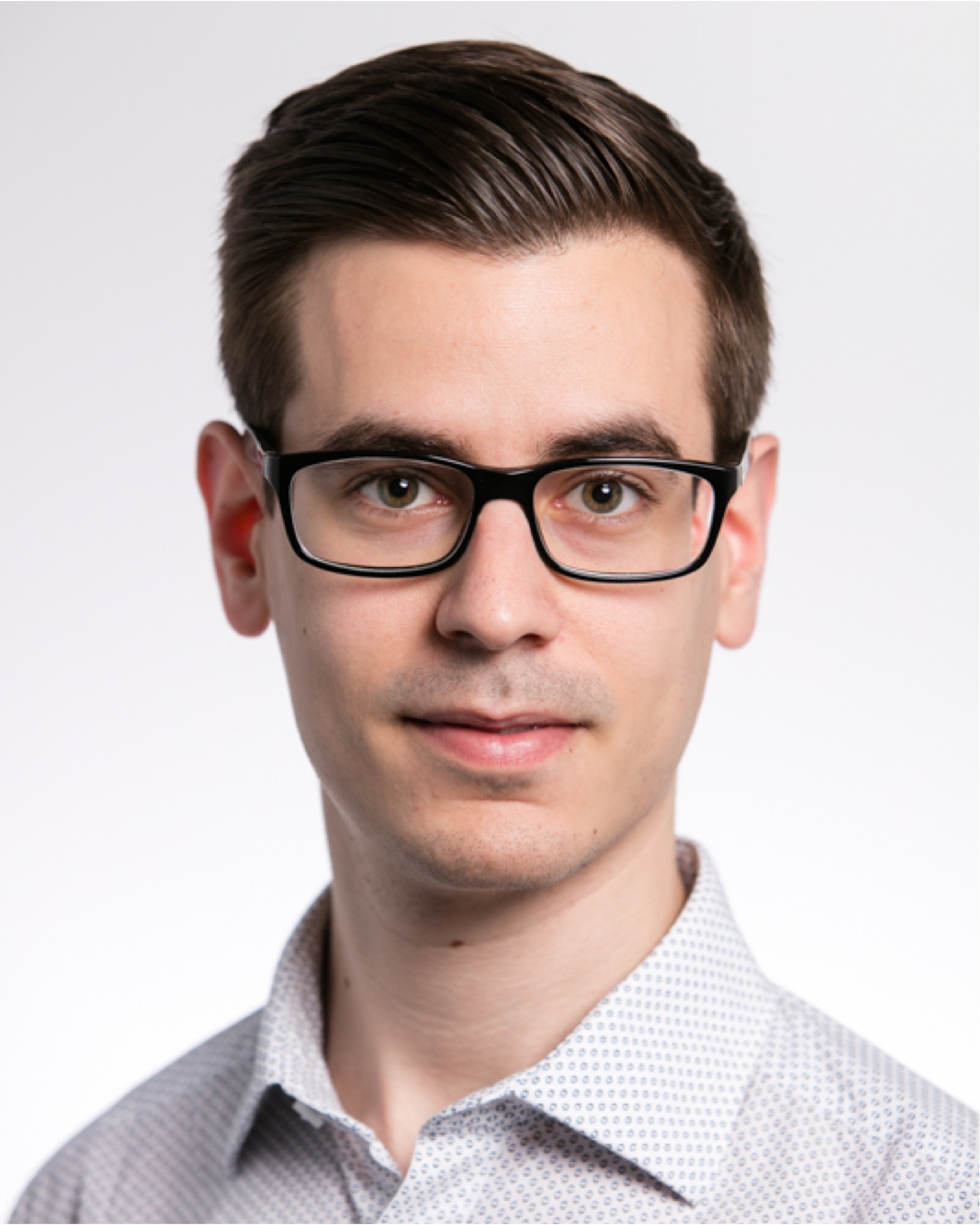}}]{Daniel Michelsanti} (S'16 M'21) received the B.Sc.\ degree in computer science and electronic engineering (cum laude) at the University of Perugia, Italy, and the M.Sc.\ degree in vision, graphics and interactive systems at Aalborg University, Denmark, in 2014 and 2017, respectively. He also received his PhD degree from Aalborg University, Denmark, in 2021. He is currently a research assistant at the section for Artificial Intelligence and Sound, Department of Electronic Systems, Aalborg University, Denmark. His research interests are in the areas of multimodal speech enhancement and machine learning, specifically deep learning.
\end{IEEEbiography}

\begin{IEEEbiography}[{\includegraphics[width=1in,height=1.25in,clip,keepaspectratio]{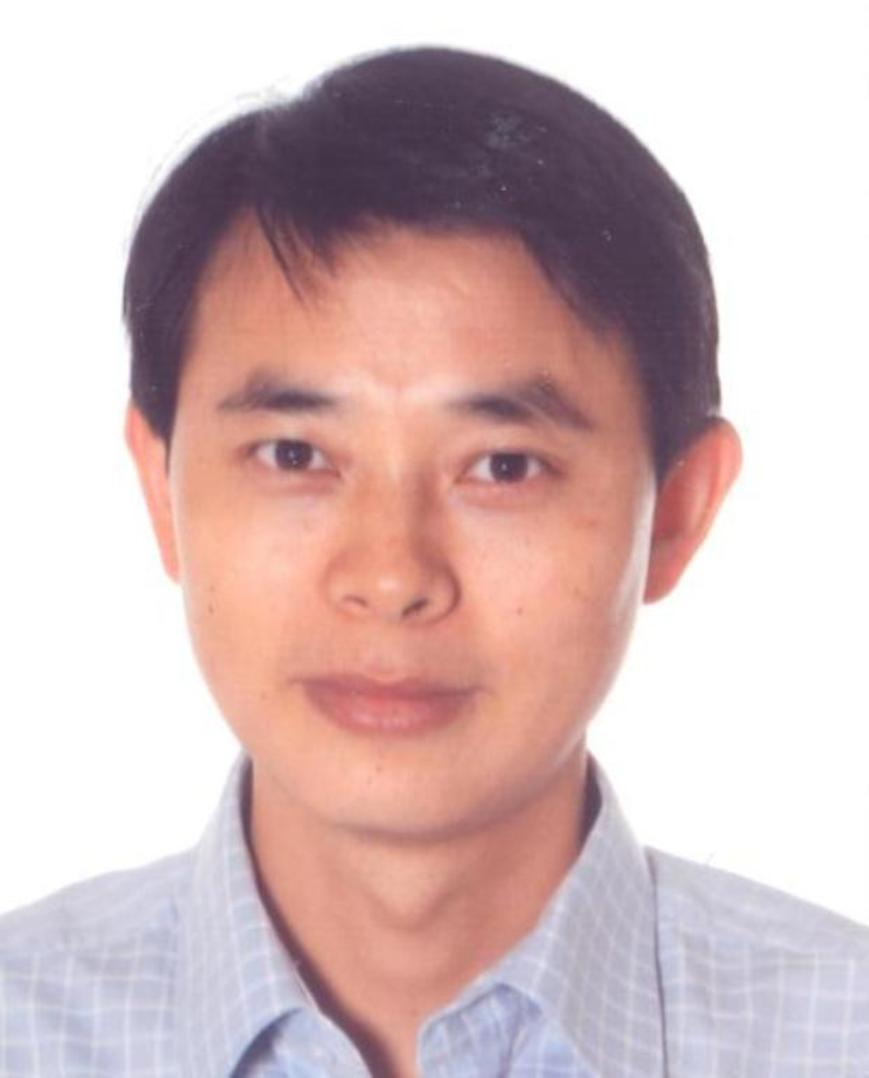}}]{Zheng-Hua Tan} (M'00 SM'06) received the B.Sc.\ and M.Sc.\ degrees in electrical engineering from Hunan University, Changsha, China, in 1990 and 1996, respectively, and the Ph.D. degree in electronic engineering from Shanghai Jiao Tong University (SJTU), Shanghai, China, in 1999.
He is a Professor in the Department of Electronic Systems and a Co-Head of the Centre for Acoustic Signal Processing Research at Aalborg University, Aalborg, Denmark. He was a Visiting Scientist at the Computer Science and Artificial Intelligence Laboratory, MIT, Cambridge, USA, an Associate Professor at SJTU, Shanghai, China, and a postdoctoral fellow at KAIST, Daejeon, Korea. His research interests include machine learning, deep learning, pattern recognition, speech and speaker recognition, noise-robust speech processing, multimodal signal processing, and social robotics. He has authored/coauthored over 200 refereed publications.
He is the Chair of the IEEE Signal Processing Society Machine Learning for Signal Processing Technical Committee (MLSP TC). He is an Associate Editor for the IEEE/ACM TRANSACTIONS ON AUDIO, SPEECH AND LANGUAGE PROCESSING. He has served as an Editorial Board Member for Computer Speech and Language and was a Guest Editor for the IEEE JOURNAL OF SELECTED TOPICS IN SIGNAL PROCESSING and Neurocomputing. He was the General Chair for IEEE MLSP 2018 and a TPC Co-Chair for IEEE SLT 2016.
\end{IEEEbiography}

\begin{IEEEbiography}[{\includegraphics[width=1in,height=1.25in,clip,keepaspectratio]{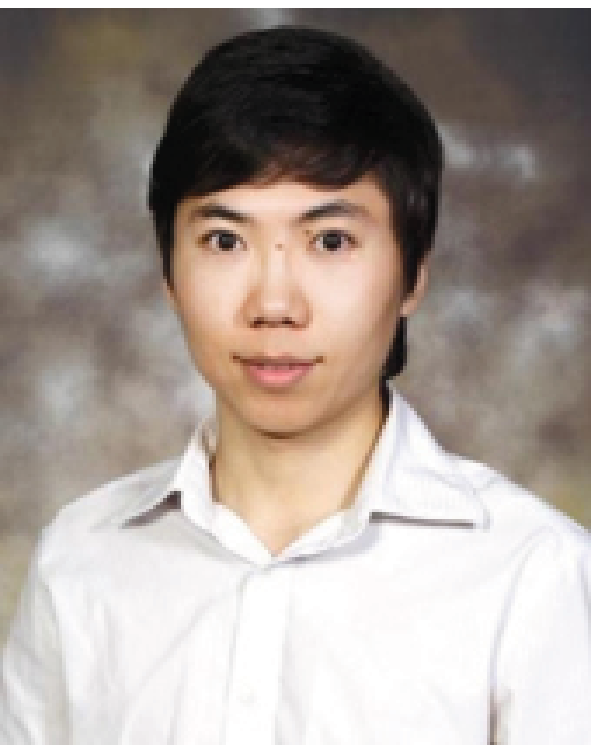}}]{Shi-Xiong Zhang}
(S'07 M'15) received the M.Phil. degree in Electronic and Information Engineering from The Hong Kong Polytechnic University in 2008 and the Ph.D. degree in the Machine Intelligence Laboratory, Engineering Department, Cambridge University in 2014. From 2014 to 2018, he was a senior speech scientist at Microsoft, speech group. Currently he is a principal researcher at Tencent America.  His research interests include speech recognition, speaker verification, speech separation, multi-modal learning and machine learning (particularly structured prediction, graphical models, kernel methods and Bayesian non-parametric methods). He was granted the ``IC Greatness award" in Microsoft in 2015. Shi-Xiong Zhang was nominated a 2011 Interspeech Best Student Paper Award for his paper ``Structured Support Vector Machines for Noise Robust Continuous Speech Recognition". He was awarded Best Paper Award in 2008 IEEE Signal Processing Postgraduate Forum for his paper ``Articulatory-Feature based Sequence Kernel For High-Level Speaker Verification".
\end{IEEEbiography}

\begin{IEEEbiography}[{\includegraphics[width=1in,height=1.25in,clip,keepaspectratio]{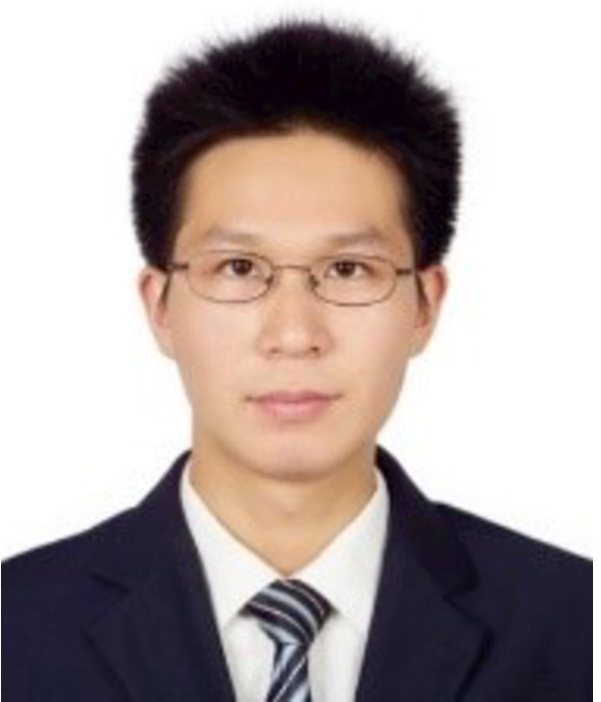}}]{Yong Xu} (M'16) received his PhD degree from University of Science and Technology of China (USTC) in 2015. He is currently a senior researcher in Tencent AI lab, Bellevue, USA. Before joining Tencent, he was a research fellow in University of Surrey, UK from April 2016 to May 2018. He once studied in Georgia Institute of Technology, USA from Sept., 2014 to May, 2015. From 2015 to 2016, he worked in iFLYTEK as a researcher. His current research interests include deep learning based speech enhancement, speech separation, noise robust speech recognition, sound event detection, etc. He received the IEEE Signal Processing Society 2018 Best Paper award for the work on deep neural networks based speech enhancement.
\end{IEEEbiography}

\vfill

\begin{IEEEbiography}[{\includegraphics[width=1in,height=1.25in,clip,keepaspectratio]{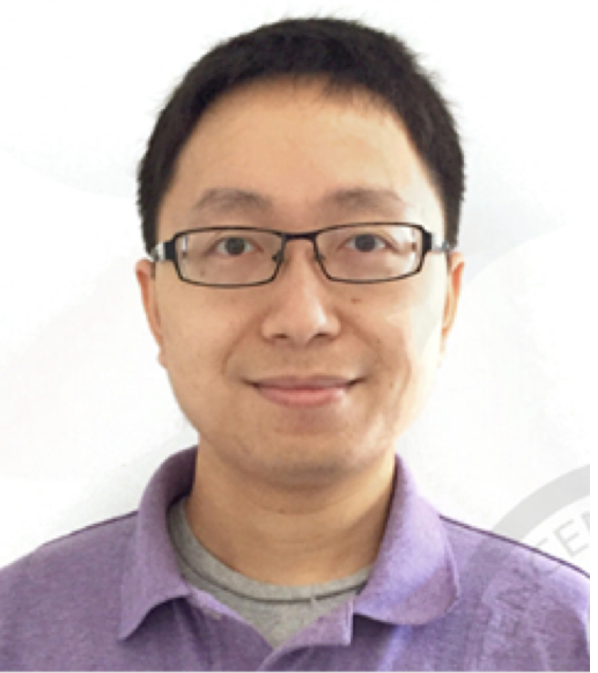}}]{Meng Yu}
received his B.S in computational mathematics at Peking University in 2007 and Ph.D in applied mathematics at University of California, Irvine in 2012. Currently he is a principal research scientist at Tencent AI Lab, working on far field frontend speech processing, deep learning based speech enhancement and separation, and their joint optimization with keyword spotting, speaker verification and acoustic model of speech recognition. 
\end{IEEEbiography}

\begin{IEEEbiography}[{\includegraphics[width=1in,height=1.25in,clip,keepaspectratio]{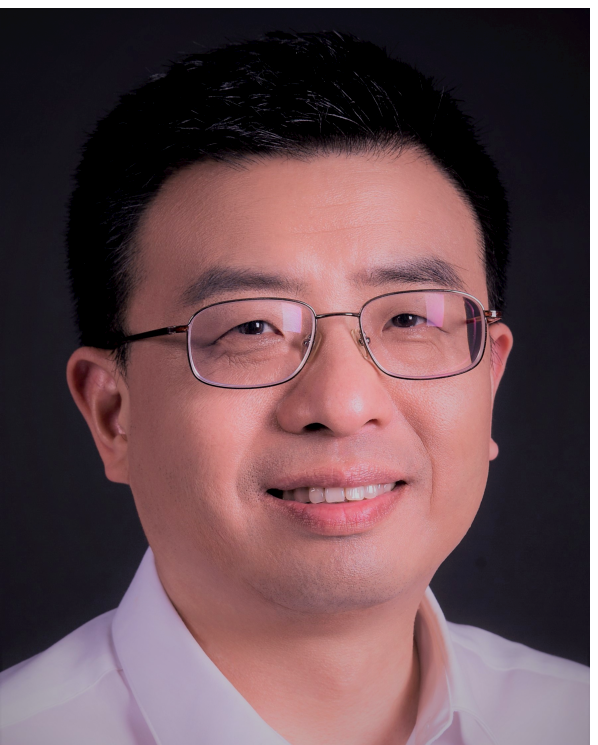}}]{Dong Yu}
(M'97 SM'06 F'18) is a distinguished scientist and vice general manager at Tencent AI Lab. Prior to joining Tencent in 2017, he was a principal researcher at Microsoft Research (Redmond), Microsoft, where he joined in 1998. He has been focusing his research on speech recognition and processing and has published two monographs and 250+ papers. His works have been cited for over 30,000 times per Google Scholar and have been recognized by the prestigious IEEE Signal Processing Society 2013 and 2016 best paper award. Dr. Dong Yu is currently serving as the vice chair of the IEEE Speech and Language Processing Technical Committee (SLPTC). He has served as a member of the IEEE SLPTC (2013-2018), a distinguished lecturer of APSIPA (2017-2018), an associate editor of the IEEE/ACM transactions on audio, speech, and language processing (TransASLP) (2011-2015), an associate editor of the IEEE signal processing magazine (2008-2011), the lead guest editor of the IEEE TransASLP - special issue on deep learning for speech and language processing (2010-2011), a guest editor of the IEEE/CAA journal of automatica sinica - special issue on deep learning in audio, image, and text processing (2015-2016), and members of organization and technical committees of many conferences and workshops.
\end{IEEEbiography}

\begin{IEEEbiography}[{\includegraphics[width=1in,height=1.25in,clip,keepaspectratio]{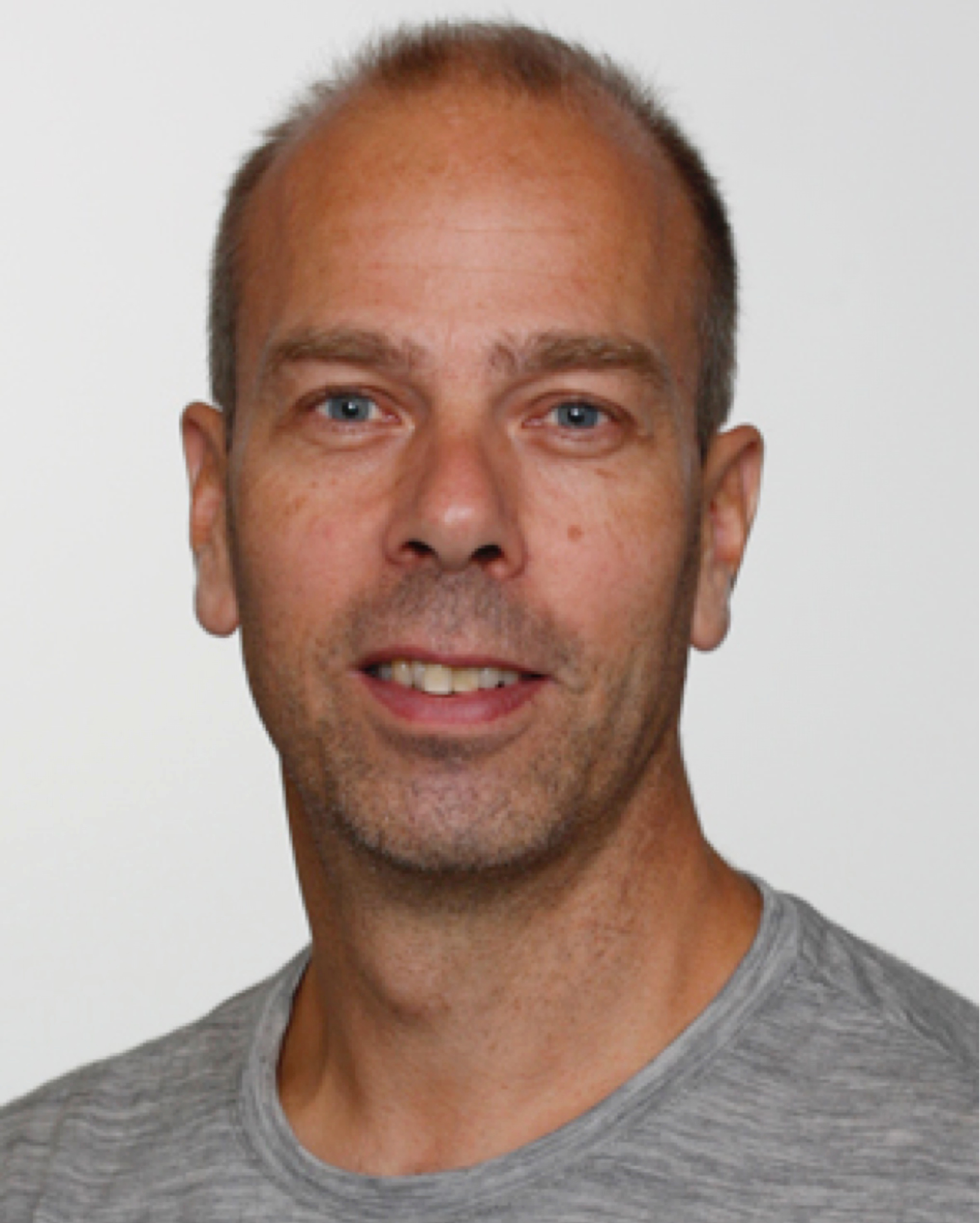}}]{Jesper Jensen}
received the M.Sc. degree in electrical
engineering and the Ph.D. degree in signal processing
from Aalborg University, Aalborg, Denmark,
in 1996 and 2000, respectively. From 1996 to 2000,
he was with the Center for Person Kommunikation
(CPK), Aalborg University, as a Ph.D. student and
Assistant Research Professor. From 2000 to 2007, he
was a Post-Doctoral Researcher and Assistant Professor
with Delft University of Technology, Delft,
The Netherlands, and an External Associate Professor
with Aalborg University. Currently, he is a Senior Principal Scientist with Oticon A/S, Copenhagen, Denmark, where his main responsibility is scouting and development of new signal processing concepts
for hearing aid applications. He is a Professor with the Section for Artificial
Intelligence and Sound (AIS), Department of Electronic Systems, at
Aalborg University. He is also a co-founder of the Centre for Acoustic Signal
Processing Research (CASPR) at Aalborg University. His main interests are
in the area of acoustic signal processing, including signal retrieval from
noisy observations, coding, speech and audio modification and synthesis,
intelligibility enhancement of speech signals, signal processing for hearing
aid applications, and perceptual aspects of signal processing.
\end{IEEEbiography}

\vfill




\end{document}